\documentclass[%
a4paper,							
11pt,								
bibliography=totoc,						
abstracton,					
]
{scrartcl}

\usepackage[a4paper,left=3.0cm,right=2.5cm, top=2.5cm, bottom=3.0cm]{geometry}
\usepackage[headsepline=.4pt,footsepline=.4pt,automark,autooneside=false,]{scrlayer-scrpage}
\clearscrheadfoot 
\automark[subsection]{section} 
\automark*[subsubsection]{subsection}
\ihead{\scriptsize\rightmark} 
\usepackage[ngerman, english]{babel}
\usepackage[T1]{fontenc}
\usepackage[utf8]{inputenc}
\usepackage{lmodern} 
\usepackage[onehalfspacing]{setspace} 

\usepackage{xspace}
\usepackage{calc}

\usepackage{amsmath}
\usepackage{amssymb}
\usepackage{amsthm,thmtools}

\usepackage{mathtools}
\usepackage{bbm}
\begingroup
    \makeatletter
    \@for\theoremstyle:=definition,remark,plain\do{%
        \expandafter\g@addto@macro\csname th@\theoremstyle\endcsname{%
            \addtolength\thm@preskip\parskip
            }%
        }
\endgroup

\usepackage{chngcntr}
\counterwithin*{equation}{section}

\theoremstyle{plain}
\newtheorem{theorem}{Theorem}[section]
\newtheorem{lemma}[theorem]{Lemma}
\theoremstyle{definition}

\theoremstyle{remark}
\newtheorem{remark}[theorem]{Remark}
\theoremstyle{plain}

\theoremstyle{plain}

\theoremstyle{remark}

\usepackage{tabularx}
\usepackage{booktabs}
\usepackage{multirow}
\usepackage{multicol}
\usepackage{pdflscape}

\usepackage{diagbox}

\usepackage[flushleft]{paralist}
\setdefaultenum{(i)}{(a)}{(1)}{(aa)}

\usepackage{color}

\usepackage{graphicx}
\usepackage{tikz}
\usetikzlibrary{calc,spy,backgrounds}
\usepackage{todonotes}
\usepackage[final]{showkeys}

\usepackage{fancyvrb}

\usepackage{float}
\usepackage[section]{placeins}
\usepackage[format=plain]{caption}

\usepackage{titleref}
\usepackage{hyperref}
\usepackage{cleveref}
\hypersetup{
    colorlinks,
		linkcolor={red},
    citecolor={blue},
    urlcolor={blue}
}

\newcommand{\R}{\mathbb{R}}

\newcommand{\Q}{\mathbb{Q}}
\newcommand{\N}{\mathbb{N}}

\newcommand{\abs}[1]{\left|#1\right|}

\newcommand{\CPU}{Intel(R) Core(TM) i7-8750H CPU @ 2.20\,GHz\xspace}
\newcommand{\RAM}{2x32\,GB (Dual Channel) Samsung SODIMM DDR4 RAM @ 2667 MHz\xspace}

\newcommand{\OS}{Windows 10 Pro\xspace}

\SaveVerb{matlab}=Matlab 2021a=
\newcommand{\matlab}{\protect\UseVerb{matlab}\xspace}
\SaveVerb{ga}=ga=
\newcommand{\ga}{\protect\UseVerb{ga}\xspace}
\SaveVerb{fmincon}=fmincon=
\newcommand{\fmincon}{\protect\UseVerb{fmincon}\xspace}
\newcommand{\matlabGOtoolbox}{(Global) Optimization Toolbox\xspace}
\newcommand{\dateA}{30/12/2019\xspace}
\newcommand{\dateB}{30/11/2020\xspace}
\newcommand{\figureSuffix}{-eps-converted-to.pdf}
\newcommand{\phiOneXA}{0.710501}
\newcommand{\phiTwoXA}{0.644564}
\newcommand{\phiThreeXA}{1.60862}
\newcommand{\xZeroA}{0.268914}
\newcommand{\phiOneYA}{0.468673}
\newcommand{\phiTwoYA}{0.533206}
\newcommand{\phiThreeYA}{1.50249}
\newcommand{\yZeroA}{0.280095}
\newcommand{\errFminA}{3.247465e-04}
\newcommand{\MREA}{0.144\,\%}

\newcommand{\phiOneXB}{0.767497}
\newcommand{\phiTwoXB}{0.699649}
\newcommand{\phiThreeXB}{1.6014}
\newcommand{\xZeroB}{0.257145}
\newcommand{\phiOneYB}{0.523363}
\newcommand{\phiTwoYB}{0.594629}
\newcommand{\phiThreeYB}{1.49966}
\newcommand{\yZeroB}{0.270007}
\newcommand{\errFminB}{3.548162e-04}
\newcommand{\MREB}{0.138\,\%}

\newcommand{\kXA}{0.578626}
\newcommand{\sigmaXA}{0.291551}
\newcommand{\thetaXA}{0.118155}
\newcommand{\kYA}{0.59774}
\newcommand{\sigmaYA}{0.262334}
\newcommand{\thetaYA}{0.0864925}

\newcommand{\kXB}{0.631802}
\newcommand{\sigmaXB}{0.308122}
\newcommand{\thetaXB}{0.120319}
\newcommand{\kYB}{0.665895}
\newcommand{\sigmaYB}{0.291125}
\newcommand{\thetaYB}{0.0954364}

\newcommand{\marketDataA}{
\renewcommand{\arraystretch}{1}
\begin{tabular}{|*{3}{c}|}
\hline
Maturity (in years) & Zero rate (in \%) & Zero-coupon price\\
\hline
$0.0833333333333333$ & $-0.469999993219972$ & $  1.0004001991529$ \\
$             0.25$ & $-0.388000020757318$ & $ 1.00096969387991$ \\
$              0.5$ & $-0.324999983422458$ & $ 1.00163343819125$ \\
$             0.75$ & $-0.314333918504417$ & $ 1.00237481461989$ \\
$                1$ & $-0.322000007145107$ & $ 1.00323926670136$ \\
$             1.25$ & $-0.323286440253412$ & $ 1.00405360258242$ \\
$              1.5$ & $-0.316161320131414$ & $ 1.00476558980205$ \\
$             1.75$ & $-0.303842297803669$ & $ 1.00535001652119$ \\
$                2$ & $-0.289547047577798$ & $ 1.00582418019158$ \\
$             2.25$ & $-0.275860329135469$ & $ 1.00623288634409$ \\
$              2.5$ & $-0.262835313503729$ & $   1.006604855007$ \\
$             2.75$ & $-0.249892233800608$ & $ 1.00691299093433$ \\
$                3$ & $-0.236451346427202$ & $ 1.00713375064174$ \\
$             3.25$ & $-0.222084053437044$ & $ 1.00725039326453$ \\
$              3.5$ & $-0.20696636298112$ & $ 1.00728054250496$ \\
$             3.75$ & $-0.191425434683623$ & $ 1.00721781901104$ \\
$                4$ & $-0.175788428168744$ & $ 1.00706740209126$ \\
$             4.25$ & $-0.160311330630236$ & $ 1.00684531811395$ \\
$              4.5$ & $-0.144965462482105$ & $ 1.00655553463348$ \\
$             4.75$ & $-0.129650957156002$ & $ 1.00618948972951$ \\
$                5$ & $-0.114267959725112$ & $ 1.00573933685071$ \\
$             5.25$ & $-0.0987154224631581$ & $ 1.00520062530541$ \\
$              5.5$ & $-0.0828875612342017$ & $ 1.00457454544122$ \\
$             5.75$ & $-0.0666773874613114$ & $ 1.00384671986489$ \\
$                6$ & $-0.0499779242090881$ & $ 1.00300667524933$ \\
$             6.25$ & $-0.0327643402378897$ & $ 1.00205088034181$ \\
$              6.5$ & $-0.0153403983915723$ & $ 1.00099833086134$ \\
$             6.75$ & $0.00190798987986796$ & $0.999871102605028$ \\
$                7$ & $0.0185949131264351$ & $0.998698306220564$ \\
$             7.25$ & $0.0344518735623467$ & $0.997505079039002$ \\
$              7.5$ & $0.0496800311054812$ & $0.996279818846146$ \\
$             7.75$ & $0.0645979575189415$ & $0.995003816465917$ \\
$                8$ & $0.0795242260210216$ & $0.993656440330286$ \\
$             8.25$ & $0.0947347900819295$ & $0.992214008696662$ \\
$              8.5$ & $0.110335148849572$ & $0.990662992494919$ \\
$             8.75$ & $0.126388167535652$ & $0.988997743889118$ \\
$                9$ & $0.142956722993404$ & $0.987213788328959$ \\
$             9.25$ & $0.160050573928316$ & $0.985308478446392$ \\
$              9.5$ & $0.177466994199449$ & $0.983284710270437$ \\
$             9.75$ & $0.194950156980411$ & $0.981173005874126$ \\
$               10$ & $0.212244223803282$ & $0.979004189945635$ \\
$               15$ & $0.473523046821356$ & $0.931543316237289$ \\
$               20$ & $0.611338950693607$ & $0.885166902653398$ \\
$               25$ & $0.652327481657267$ & $0.849865688031976$ \\
$               30$ & $0.640345783904195$ & $0.825611308910539$ \\\hline
\end{tabular}
}

\newcommand{\marketDataB}{
\renewcommand{\arraystretch}{1}
\begin{tabular}{|*{3}{c}|}
\hline
Maturity (in years) & Zero rate (in \%) & Zero-coupon price\\
\hline
$0.0833333333333333$ & $-0.499999988824129$ & $ 1.00041207460911$ \\
$0.244444444444444$ & $-0.526142632588744$ & $ 1.00130160930003$ \\
$              0.5$ & $-0.507755391299725$ & $ 1.00252751322004$ \\
$             0.75$ & $-0.503638433292508$ & $ 1.00378359716281$ \\
$                1$ & $-0.517199980095029$ & $ 1.00519888845098$ \\
$ 1.24444444444444$ & $-0.524928161568994$ & $ 1.00658242962185$ \\
$              1.5$ & $-0.525975602238304$ & $ 1.00791998180754$ \\
$             1.75$ & $-0.522984338103072$ & $ 1.00920759910672$ \\
$                2$ & $-0.518596358597279$ & $ 1.01045317135578$ \\
$ 2.24444444444444$ & $-0.514924664329897$ & $ 1.01166557509739$ \\
$              2.5$ & $-0.511966253088758$ & $  1.0128933501334$ \\
$             2.75$ & $-0.509189122195153$ & $ 1.01412683679008$ \\
$                3$ & $-0.50606126897037$ & $ 1.01533680279481$ \\
$ 3.24722222222222$ & $-0.502153361016866$ & $  1.0164921683478$ \\
$              3.5$ & $-0.497446840398652$ & $ 1.01760034898714$ \\
$             3.75$ & $-0.492025761253245$ & $ 1.01867206863544$ \\
$                4$ & $-0.485974224284291$ & $ 1.01969106459461$ \\
$ 4.24444444444444$ & $-0.479376664922526$ & $ 1.02062910472473$ \\
$              4.5$ & $-0.47231881015648$ & $ 1.02152663951854$ \\
$             4.75$ & $-0.464886791550967$ & $ 1.02238363928715$ \\
$                5$ & $-0.457166694104671$ & $ 1.02318805666289$ \\
$ 5.24444444444444$ & $-0.449221897941854$ & $ 1.02391569856637$ \\
$              5.5$ & $-0.441024916645461$ & $  1.0246017102629$ \\
$             5.75$ & $-0.432525547282481$ & $ 1.02524040701367$ \\
$                6$ & $-0.423673586919904$ & $ 1.02581359580938$ \\
$ 6.24444444444444$ & $-0.414435069675712$ & $ 1.02629288605375$ \\
$              6.5$ & $-0.404840884313273$ & $ 1.02671307570551$ \\
$             6.75$ & $-0.394938214854612$ & $ 1.02707383363012$ \\
$                7$ & $-0.384774198755622$ & $ 1.02736441987847$ \\
$ 7.24722222222222$ & $-0.37439006592308$ & $ 1.02757527517123$ \\
$              7.5$ & $-0.363803462582268$ & $ 1.02771717281163$ \\
$             7.75$ & $-0.353026057560157$ & $  1.0277994267812$ \\
$                8$ & $-0.342069566249847$ & $ 1.02781095336332$ \\
$ 8.24444444444444$ & $-0.330958580803831$ & $ 1.02773379298185$ \\
$              8.5$ & $-0.319769247063562$ & $ 1.02760240180053$ \\
$             8.75$ & $-0.308590599271419$ & $ 1.02742329257773$ \\
$                9$ & $-0.297511671669781$ & $ 1.02719551414522$ \\
$ 9.24444444444444$ & $-0.286599750659389$ & $ 1.02691014728746$ \\
$              9.5$ & $-0.275835084599585$ & $ 1.02659173861864$ \\
$             9.75$ & $-0.265176203111905$ & $ 1.02623638422587$ \\
$               10$ & $-0.254581612534821$ & $ 1.02583261454546$ \\
$               15$ & $-0.0622837862465531$ & $ 1.00939445284171$ \\
$               20$ & $0.0184025324415416$ & $0.996324084241296$ \\
$               25$ & $0.0234601888223551$ & $0.994148968990786$ \\
$               30$ & $-0.00393075206375215$ & $ 1.00118069913068$ \\\hline
\end{tabular}
}

\newcommand{\dfMeanErrABody}{
$0.0833333$ & $0.000409963$ \\
$0.25$ & $0.00112268$ \\
$0.5$ & $0.00161849$ \\
$0.75$ & $0.00137477$ \\
$1$ & $0.000814101$ \\
$1.25$ & $0.000328491$ \\
$1.5$ & $1.4362e-05$ \\
$1.75$ & $5.62967e-05$ \\
$2$ & $0.000402274$ \\
$2.25$ & $0.000824884$ \\
$2.5$ & $0.0011063$ \\
$2.75$ & $0.00133005$ \\
$3$ & $0.00143482$ \\
$3.25$ & $0.0018223$ \\
$3.5$ & $0.00242705$ \\
$3.75$ & $0.00303645$ \\
$4$ & $0.00354305$ \\
$4.25$ & $0.00409665$ \\
$4.5$ & $0.00461513$ \\
$4.75$ & $0.00491301$ \\
$5$ & $0.00502117$ \\
$5.25$ & $0.00485501$ \\
$5.5$ & $0.0048752$ \\
$5.75$ & $0.00549422$ \\
$6$ & $0.00642713$ \\
$6.25$ & $0.00743045$ \\
$6.5$ & $0.00841358$ \\
$6.75$ & $0.00945159$ \\
$7$ & $0.010154$ \\
$7.25$ & $0.0105884$ \\
$7.5$ & $0.0111652$ \\
$7.75$ & $0.0116295$ \\
$8$ & $0.0120698$ \\
$8.25$ & $0.0125788$ \\
$8.5$ & $0.0132799$ \\
$8.75$ & $0.0138392$ \\
$9$ & $0.0146873$ \\
$9.25$ & $0.0156794$ \\
$9.5$ & $0.0166502$ \\
$9.75$ & $0.0174248$ \\
$10$ & $0.0180652$ \\
$15$ & $0.0251373$ \\
$20$ & $0.0331911$ \\
$25$ & $0.0160001$ \\
$30$ & $0.00094742$ 
}

\newcommand{\dfMeanErrBData}{
$0.000560815$ \\
$0.00119113$ \\
$0.00144972$ \\
$0.00113393$ \\
$0.000550498$ \\
$0.000398677$ \\
$0.000571146$ \\
$0.000805009$ \\
$0.00125292$ \\
$0.00191315$ \\
$0.00288173$ \\
$0.00356734$ \\
$0.00401817$ \\
$0.00441028$ \\
$0.00477583$ \\
$0.00512345$ \\
$0.00537247$ \\
$0.0051572$ \\
$0.00473785$ \\
$0.00449665$ \\
$0.00423046$ \\
$0.00398605$ \\
$0.00410309$ \\
$0.00442951$ \\
$0.00477755$ \\
$0.00495872$ \\
$0.00514248$ \\
$0.00546137$ \\
$0.00596334$ \\
$0.006623$ \\
$0.00753078$ \\
$0.00803137$ \\
$0.00842589$ \\
$0.00889094$ \\
$0.00938679$ \\
$0.00983092$ \\
$0.0100521$ \\
$0.0103367$ \\
$0.0109601$ \\
$0.0117957$ \\
$0.0126498$ \\
$0.0314676$ \\
$0.0233781$ \\
$0.0174282$ \\
$0.00846645$ 
}

\newcommand{\simulationsA}{10000}
\newcommand{\timeMeshA}{\frac{1}{256}}

\newcommand{\errorSwaptionA}{
\renewcommand{\arraystretch}{1}
\begin{tabular}{|c|*{5}{c}|}
\hline
\diagbox{Maturity}{Tenor} & 1 & 2 & 5 & 7 & 10 \\
 \hline
$1$ & $4.82602\,\% $ & $3.72741\,\% $ & $1.56009\,\% $ & $0.601718\,\% $ & $-0.554628\,\% $ \\
$2$ & $4.70047\,\% $ & $3.56775\,\% $ & $1.18057\,\% $ & $0.0467543\,\% $ & $-1.37405\,\% $ \\
$5$ & $3.60511\,\% $ & $2.4125\,\% $ & $-0.249531\,\% $ & $-1.64559\,\% $ & $-3.50471\,\% $ \\
$7$ & $3.23362\,\% $ & $1.97696\,\% $ & $-0.945175\,\% $ & $-2.48803\,\% $ & $-4.54551\,\% $ \\
$10$ & $2.85895\,\% $ & $1.49467\,\% $ & $-1.69265\,\% $ & $-3.37829\,\% $ & $-5.63478\,\% $ \\
$15$ & $2.63259\,\% $ & $1.17927\,\% $ & $-2.21714\,\% $ & $-4.05043\,\% $ & $-6.53224\,\% $ \\
$20$ & $2.48029\,\% $ & $1.01554\,\% $ & $-2.47166\,\% $ & $-4.39298\,\% $ & $-6.98019\,\% $ 
\\\hline
\end{tabular}
}

\newcommand{\errorSwaptionB}{
\renewcommand{\arraystretch}{1}
\begin{tabular}{|c|*{5}{c}|}
\hline
\diagbox{Maturity}{Tenor} & 1 & 2 & 5 & 7 & 10 \\
 \hline
$1$ & $4.87829\,\% $ & $3.71571\,\% $ & $1.62322\,\% $ & $0.745822\,\% $ & $-0.307305\,\% $ \\
$2$ & $4.62191\,\% $ & $3.4627\,\% $ & $1.22337\,\% $ & $0.167669\,\% $ & $-1.18931\,\% $ \\
$5$ & $3.82762\,\% $ & $2.596\,\% $ & $-0.0624237\,\% $ & $-1.4526\,\% $ & $-3.34358\,\% $ \\
$7$ & $3.45837\,\% $ & $2.13705\,\% $ & $-0.80375\,\% $ & $-2.35496\,\% $ & $-4.46368\,\% $ \\
$10$ & $3.24577\,\% $ & $1.77121\,\% $ & $-1.56541\,\% $ & $-3.3177\,\% $ & $-5.74088\,\% $ \\
$15$ & $3.02788\,\% $ & $1.4328\,\% $ & $-2.26341\,\% $ & $-4.2787\,\% $ & $-7.03714\,\% $ \\
$20$ & $3.16548\,\% $ & $1.42331\,\% $ & $-2.6344\,\% $ & $-4.8421\,\% $ & $-7.85129\,\% $ 
\\\hline
\end{tabular}
}

\newcommand{\marketSwaptionA}{
\renewcommand{\arraystretch}{1}
\begin{tabular}{|c|*{5}{c}|}
\hline
\diagbox{Maturity}{Tenor} & 1 & 2 & 5 & 7 & 10 \\
 \hline
$1$ & $0.000702236$ & $0.00175071$ & $0.00706456$ & $0.0112631$ & $0.0181169$ \\
$2$ & $0.0014433$ & $0.00333027$ & $0.0111956$ & $0.017189$ & $0.0265694$ \\
$5$ & $0.00391314$ & $0.00796766$ & $0.0214221$ & $0.0308074$ & $0.0453508$ \\
$7$ & $0.00521117$ & $0.0104082$ & $0.0268942$ & $0.0380283$ & $0.0548627$ \\
$10$ & $0.00668368$ & $0.0132567$ & $0.0330802$ & $0.045932$ & $0.0651091$ \\
$15$ & $0.00781681$ & $0.0154396$ & $0.0378811$ & $0.0525334$ & $0.0743464$ \\
$20$ & $0.00840243$ & $0.0166069$ & $0.0407885$ & $0.0565876$ & $0.0795953$ 
\\\hline
\end{tabular}
}

\newcommand{\marketSwaptionB}{
\renewcommand{\arraystretch}{1}
\begin{tabular}{|c|*{5}{c}|}
\hline
\diagbox{Maturity}{Tenor} & 1 & 2 & 5 & 7 & 10 \\
 \hline
$1$ & $0.000661505$ & $0.00151259$ & $0.00547626$ & $0.00900521$ & $0.0149394$ \\
$2$ & $0.00121578$ & $0.00278226$ & $0.00916199$ & $0.0145571$ & $0.0235389$ \\
$5$ & $0.00336776$ & $0.00707345$ & $0.0194074$ & $0.0285692$ & $0.0433793$ \\
$7$ & $0.0047279$ & $0.00963075$ & $0.0253482$ & $0.0364087$ & $0.0537919$ \\
$10$ & $0.00630168$ & $0.0126162$ & $0.0323949$ & $0.0456958$ & $0.0665005$ \\
$15$ & $0.00790807$ & $0.0157371$ & $0.0397201$ & $0.0558865$ & $0.0802858$ \\
$20$ & $0.00898181$ & $0.017927$ & $0.0451252$ & $0.0631302$ & $0.0898938$ 
\\\hline
\end{tabular}
}

\newcommand{\strikeSwaptionA}{
\renewcommand{\arraystretch}{1}
\begin{tabular}{|c|*{5}{c}|}
\hline
\diagbox{Maturity}{Tenor} & 1 & 2 & 5 & 7 & 10 \\
 \hline
$1$ & $-0.260793\,\% $ & $-0.195187\,\% $ & $-0.011405\,\% $ & $0.140129\,\% $ & $0.330514\,\% $ \\
$2$ & $-0.129665\,\% $ & $-0.0782444\,\% $ & $0.139932\,\% $ & $0.273273\,\% $ & $0.449172\,\% $ \\
$5$ & $0.268095\,\% $ & $0.38307\,\% $ & $0.556996\,\% $ & $0.655339\,\% $ & $0.757978\,\% $ \\
$7$ & $0.547079\,\% $ & $0.611571\,\% $ & $0.76683\,\% $ & $0.830788\,\% $ & $0.891069\,\% $ \\
$10$ & $0.880582\,\% $ & $0.907944\,\% $ & $0.967521\,\% $ & $0.988131\,\% $ & $0.992003\,\% $ \\
$15$ & $1.04232\,\% $ & $1.04153\,\% $ & $1.01776\,\% $ & $0.985317\,\% $ & $0.924744\,\% $ \\
$20$ & $0.925377\,\% $ & $0.901441\,\% $ & $0.827386\,\% $ & $0.778437\,\% $ & $0.721445\,\% $ 
\\\hline
\end{tabular}
}

\newcommand{\strikeSwaptionB}{
\renewcommand{\arraystretch}{1}
\begin{tabular}{|c|*{5}{c}|}
\hline
\diagbox{Maturity}{Tenor} & 1 & 2 & 5 & 7 & 10 \\
 \hline
$1$ & $-0.558066\,\% $ & $-0.544838\,\% $ & $-0.455765\,\% $ & $-0.37221\,\% $ & $-0.238803\,\% $ \\
$2$ & $-0.531679\,\% $ & $-0.502856\,\% $ & $-0.386908\,\% $ & $-0.294521\,\% $ & $-0.162606\,\% $ \\
$5$ & $-0.315638\,\% $ & $-0.264729\,\% $ & $-0.117094\,\% $ & $-0.0324645\,\% $ & $0.0536401\,\% $ \\
$7$ & $-0.117189\,\% $ & $-0.0652544\,\% $ & $0.0603589\,\% $ & $0.1157\,\% $ & $0.150538\,\% $ \\
$10$ & $0.150213\,\% $ & $0.179568\,\% $ & $0.225372\,\% $ & $0.223805\,\% $ & $0.196761\,\% $ \\
$15$ & $0.234862\,\% $ & $0.219855\,\% $ & $0.16784\,\% $ & $0.12791\,\% $ & $0.0641018\,\% $ \\
$20$ & $0.0500327\,\% $ & $0.0277677\,\% $ & $-0.0398808\,\% $ & $-0.0837531\,\% $ & $-0.134806\,\% $ 
\\\hline
\end{tabular}
}

\newcommand{\ctimeTestsABody}{
\UseVerb{ga} &$42.136$ &$0.142014\,\%$\\
0.50001     0.50001         1.5     0.50001     0.50001         1.5     0.50001     0.50001 &$0.287$ &$0.143798\,\%$\\
1  1  2  1  1  2  1  1 &$0.229$ &$0.146769\,\%$\\
1e-05       1e-05           1       1e-05       1e-05           1       1e-05       1e-05 &$0.276$ &$0.145207\,\%$\\
0.048808     0.72079      1.4275     0.64469     0.32152      1.4794      0.2556     0.25427 &$0.334$ &$0.146509\,\%$\\
0.66073     0.78158      1.5547      0.3779     0.24209       1.249     0.74017     0.63334 &$103.949$ &$0.14374\,\%$
}
\newcommand{\ctimeTestsAHeader}{
Inital Parameter & Times (in s) & MRE (in \%) \\
}

\newcommand{\ctimeTestsBBody}{
\UseVerb{ga} &$35.842$ &$0.135885\,\%$\\
0.50001     0.50001         1.5     0.50001     0.50001         1.5     0.50001     0.50001 &$0.280$ &$0.137577\,\%$\\
1  1  2  1  1  2  1  1 &$0.188$ &$0.138228\,\%$\\
1e-05       1e-05           1       1e-05       1e-05           1       1e-05       1e-05 &$40.864$ &$0.13642\,\%$\\
0.87647     0.89591      1.4508     0.10496     0.63629      1.3009      0.3297     0.74067 &$0.185$ &$0.140284\,\%$\\
0.51541     0.83366      1.9624     0.79757     0.13068      1.6517     0.54064     0.43674 &$0.342$ &$0.14292\,\%$
}

\newcommand{\rootFigFolderA}{TermStructureEUR_20191230}
\newcommand{\rootFigFolderB}{TermStructureEUR_20201130}
\newcommand{\figRatesOneA}{\rootFigFolderA/Rates/R_A_1\figureSuffix}

\newcommand{\figMeanVarOneA}{\rootFigFolderA/MeanVar/MV_A_1\figureSuffix}
\newcommand{\figMeanVarOneB}{\rootFigFolderB/MeanVar/MV_B_1\figureSuffix}

\newcommand{\figMeanVarThreeA}{\rootFigFolderA/MeanVar/MV_A_3\figureSuffix}
\newcommand{\figMeanVarThreeB}{\rootFigFolderB/MeanVar/MV_B_3\figureSuffix}
\newcommand{\figZCOneA}{\rootFigFolderA/ZC/ZC_A_1\figureSuffix}

\newcommand{\figZCTwoA}{\rootFigFolderA/ZC/ZC_A_2\figureSuffix}

\newcommand{\figDFOneA}{\rootFigFolderA/DF/DF_A_1\figureSuffix}

\newcommand{\figDFTwoA}{\rootFigFolderA/DF/DF_A_2\figureSuffix}

\newcommand{\figFOneA}{\rootFigFolderA/Forward/F_A_1\figureSuffix}
\newcommand{\figFOneB}{\rootFigFolderB/Forward/F_B_1\figureSuffix}
\newcommand{\figFTwoA}{\rootFigFolderA/Forward/F_A_2\figureSuffix}
\newcommand{\figFTwoB}{\rootFigFolderB/Forward/F_B_2\figureSuffix}
\newcommand{\figFThreeA}{\rootFigFolderA/Forward/F_A_3\figureSuffix}
\newcommand{\figFThreeB}{\rootFigFolderB/Forward/F_B_3\figureSuffix}
\newcommand{\headlogo}{
\vspace{-5\headheight}
\raisebox{2pt}{
\includegraphics[width=.1\textwidth]{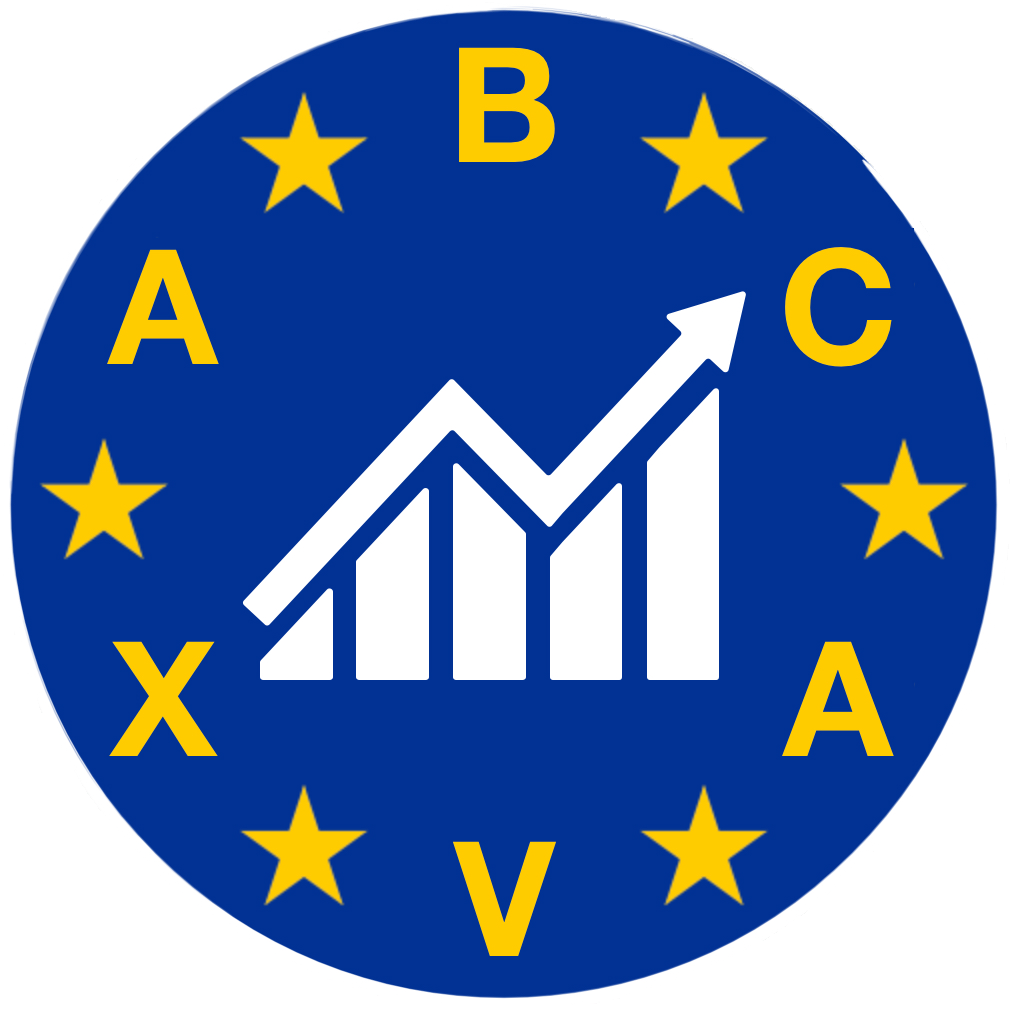}
}
\hfill
\mbox{
\includegraphics[trim=5.1cm 11cm 5.1cm 12cm,clip, width=.2\textwidth]{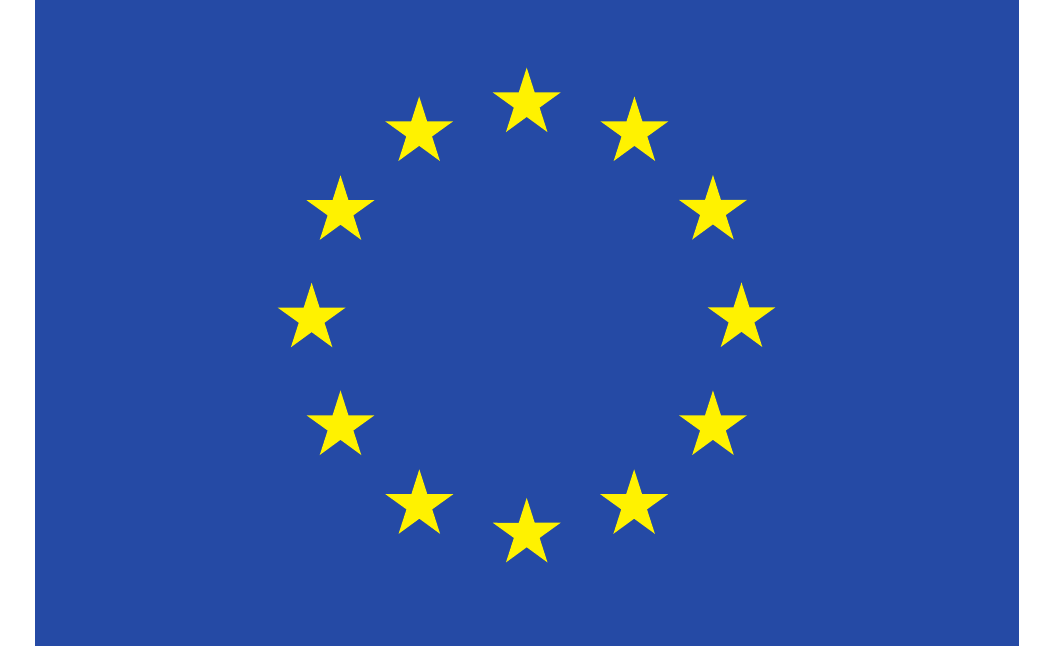}
} \hspace{-1.5em}}
\title{{\headlogo}\\
How to handle negative interest rates in a CIR framework}
\author{Marco Di Francesco\thanks{UnipolSai Assicurazioni, via Stalingrado 45, Bologna, Italy, \textbf{e-mail}: marco.difrancesco@unipolsai.com}\and Kevin Kamm\thanks{Dipartimento di Matematica, Universit\`a di Bologna, Bologna, Italy.
\textbf{e-mail}: kevin.kamm@unibo.it}
}

\begin{document}
\thispagestyle{empty}\pagenumbering{roman}
\maketitle
\renewcommand{\thefootnote}{\Roman{footnote}}
\footnotetext[0]{The views expressed in this note are the only responsibility of the author and do
not represent in any way those of author's current employer. All errors are the only
responsibility of the author.}
\renewcommand{\thefootnote}{\arabic{footnote}}
\begin{abstract}
In this paper, we propose a new model to address the problem of negative interest rates that preserves the analytical tractability of the original Cox-Ingersoll-Ross (CIR) model without introducing a shift to the market interest rates, because it is defined as the difference of two independent CIR processes. The strength of our model lies within the fact that it is very simple and can be calibrated to the market zero yield curve using an analytical formula. We run several numerical experiments at two different dates, once with a partially sub-zero interest rate and once with a fully negative interest rate. In both cases, we obtain good results in the sense that the model reproduces the market term structures very well. We then simulate the model using the Euler-Maruyama scheme and examine the mean, variance and distribution of the model. The latter agrees with the skewness and fat tail seen in the original CIR model. In addition, we compare the model's zero coupon prices with market prices at different future points in time. Finally, we test the market consistency of the model by evaluating swaptions with different tenors and maturities.  
\end{abstract}
\noindent
\textbf{Acknowledgements:}
This project has received funding from the European Union’s Horizon 2020 research and innovation programme
under the Marie Sklodowska-Curie grant agreement No 813261 and is part of the ABC-EU-XVA
project.
\newpage
\pagestyle{scrheadings}\ihead{\scriptsize\rightmark}\pagenumbering{arabic}
			%
			%
			%
			%
\section{Introduction}\label{sec:introduction}
The Cox-Ingersoll-Ross model (hereafter referred to as CIR model)
has been regarded as the reference model in interest rate modeling by both practitioners and academics for several decades, not only because of its analytical tractability as an affine model, but also because of its derivation from a general equilibrium framework (see for example \cite{CIR1981}), among other reasons. The well-known feature of the CIR model that ultimately led to this paper is that interest rates never become negative. This long-standing paradigm of non-negative interest rates made the CIR model and its extensions one of the most appropriate models for interest rate modeling.

Today, however, negative interest rates are very common and thus the need for models that can handle this paradigm shift is highly desirable, provided that they have as few shortcomings as possible compared to the original CIR models.

In this paper, we present a very simple and effective idea how this can be realized by modeling interest rates as the difference of two independent CIR processes, which---to the best of our knowledge---has not been considered yet.

We will propose a term structure in the risk-neutral world suitable for the difference of two independent affine processes and obtain a pricing formula for default-free zero-coupon bonds by deriving the associated Riccati equations arising from this no-arbitrage framework. In the special case of two CIR processes we will then solve the Riccati equations explicitly, which preserves the analytical tractability of its non-negative interest rate counterpart.

Afterwards, we will show some numerical experiments to demonstrate the merits of this approach in practice.

Let us consider the following affine dynamics
\begin{align}
	&\left\{
	\begin{aligned}[c]\arraycolsep=0pt
		dx(t)&=\left(\lambda_x(t) x(t) + \eta_x(t)\right)dt + 
			\sqrt{\gamma_x(t) x(t) + \delta_x(t)}dW_x(t)\\
		x(0)&=x_0,
	\end{aligned}
	\right.\label{eq:x}\\
	&\left\{
	\begin{aligned}[c]\arraycolsep=0pt
		dy(t)&=\left(\lambda_y(t) y(t) + \eta_y(t)\right)dt + 
			\sqrt{\gamma_y(t) y(t) + \delta_y(t)}dW_y(t)\\
		y(0)&=y_0,
	\end{aligned}
	\right.\label{eq:y}
\end{align}
where henceforth throughout the whole paper $W_y$ and $W_x$ are two independent standard Brownian motions on a stochastic basis 
$\left(\Omega,\mathcal{F},\left(\mathcal{F}_t\right)_{t\in [0,T]},\Q\right)$, $\Q$ is a martingale measure and $T>0$ is a finite time horizon. The initial values $x_0,\ y_0\in \R$ are real-valued constants and the coefficients $\lambda_z,\eta_z,\gamma_z,\delta_z$, $z\in\left\{x,y\right\}$, are all real-valued deterministic functions, such that \eqref{eq:x} and \eqref{eq:y} are well-defined.

Furthermore, let the the instantaneous short-rate process be given by
\begin{align}
	\label{eq:r}
	r(t)\coloneqq x(t)-y(t).
\end{align}

In the case where $y\equiv 0$, this reduces to the standard affine one-factor short rate model class. If additionally $\delta_x(t)\equiv 0$, $\lambda_x(t)\equiv -k_x$\,
$\eta_x(t)\equiv k_x\theta_x$ and $\gamma_x(t)\equiv \sigma_x^2$, where $k_x,\sigma_x,\theta_x\in\R_{\geq 0}$, it reduces to the standard CIR model
\begin{align}
 dx(t) = k_x(\theta_x - x(t))dt + \sigma_x \sqrt{x(t)} dW_x(t),
\label{eq:CIR}
\end{align}
which lets \eqref{eq:r} preserve all the features of a standard CIR model in a non-negative interest rate setting.
\subsection{Description of the main results}\label{sec:intro_description}
The main result consists of two main parts. First of all, we derive the zero-coupon bond price
for \eqref{eq:r} in the case of the difference of \eqref{eq:x} and \eqref{eq:y} being two independent CIR processes as in \eqref{eq:CIR}. Secondly, we provide numerical experiments to demonstrate the features of this model in \Cref{sec:numerics}.
\begin{theorem}\label{thm:main}%
	Let $\left(\Omega,\mathcal{F},\left(\mathcal{F}_t\right)_{t\in [0,T]},\Q\right)$ be a 
	stochastic basis, where $\Q$ is a martingale measure, $T>0$ a finite time horizon and let the $\sigma$-algebra $\left(\mathcal{F}_t\right)_{t\in [0,T]}$ fulfill the usual conditions and support two independent standard Brownian motions $W_x$ and $W_y$.
	
	The price of a zero-coupon bond in the model $r(t)=x(t)-y(t)$ 
	with $x$ and $y$ being two independent CIR processes as in \eqref{eq:CIR} is given by
	\begin{align}
		P(t,T) = A_x(t,T)e^{-B_x(t,T)x(t)}A_y(t,T)e^{B_y(t,T)y(t)},
		\label{eq:termstructure}
	\end{align}
	where $t\leq T$ and for $z\in \left\{x,y\right\}$
	\begin{align}\label{eq:AeB}
	\begin{aligned}[c]\arraycolsep=0pt
		A_z(t,T) &=  
			\left( 
				\frac{
					\phi^z_1 e^{\phi^z_2 (T-t)}
				}{
					\phi^z_2
						\left(
							e^{\phi^z_1(T-t)} -1
						\right)+ 
					\phi^z_1
				} 
			\right)^{\phi^z_3} \\
		B_z(t,T) &=  
			\frac{
				e^{\phi^z_1(T-t)}-1
			}{
				\phi^z_2 
					\left(
						e^{\phi^z_1(T-t)}-1
					\right)+ \phi^z_1
			}
	\end{aligned}
	\end{align} 
	with $\phi_i^z \geq 0$, 
	$i=1,2,3$, $z\in \left\{x,y\right\}$, such that the 
	Feller condition $2k_z\theta_z \geq \sigma_z^2$ is satisfied and
	\begin{align}\label{eq:phi}
	\begin{aligned}[c]
	&\phi^x_1 = \sqrt{k_x^2 + 2\sigma_x^2}, &&& 
	&\phi^x_2 = \frac{k_x + \phi^x_1}{2},&&&
	&\phi^x_3 = \frac{2k_x\theta_x}{\sigma_x^2} \\
	&\phi^y_1 = \sqrt{k_y^2 - 2\sigma_y^2},&&&
	&\phi^y_2 = \frac{k_y + \phi^y_1}{2}, &&&
	&\phi^y_3 = \frac{2k_y\theta_y}{\sigma_y^2}.
	\end{aligned}
	\end{align}
\end{theorem}
The technical part of the proof is quite standard and referred to \Cref{sec:Riccati} with a description for deriving this result in \Cref{sec:model}. Formula \eqref{eq:termstructure} will provide the necessary ingredient for the numerical experiments in \Cref{sec:numerics} to calibrate the model to the market term structure. In \Cref{sec:numerics} we will perform several experiments at two different dates \dateA and \dateB, where negative interest rates are observed in the market to uncover the features of this model.
\subsection{Review of the literature and comparison}\label{sec:review}
There is a vast literature on interest rate modeling, among these for example the comprehensive works of \cite{Bjork}, \cite{BrigoMercurioLibro} and \cite{HullLibro}, which we cannot cover to its full extent in this small review. 
The most popular approach in modern interest rate modeling is the direct modeling of short rates $r(t)$ under a risk-neutral measure $\Q$ inspired by no-arbitrage arguments, especially, because the price at time $t>0$ of a contingent claim with payoff $H_T$, $T>t$, under the risk-neutral measure is given by (cf. \cite{PascucciBook})
\begin{align}\label{eq:riskneutralPrice}
 H_t= E_t^\Q [e^{-\int_{t}^{T}r(s)ds} H_T],
\end{align}
where $E_t^\Q$ denotes the conditional expectation with respect to some filtration $\mathcal{F}_t$ under measure $\Q$. In particular, choosing $H_T\coloneqq P(T,T)=1$, where $P(t,T)$ denotes a zero-coupon bond, gives rise to a convenient way to calibrate a short rate model to the market term structure, which we will utilize for our approach as well.

Starting with the pioneering works of Merton \cite{Merton} in 1973 and Vasicek \cite{Vasicek} in 1977, many one-factor short rate models were introduced, see \cite{BrigoMercurioLibro} for a detailed overview. Among all, a model that has had a particular importance in the past, equally among both practitioners and academics, is the well-known CIR model, proposed by Cox, Ingersoll \& Ross in \cite{CIR}. It provides the basis for this paper and is a generalization to the Vasicek model by introducing a non-constant volatility given by \eqref{eq:CIR}.

Clearly, the square root term precludes the possibility of negative interest rates and under the assumption of the Feller-condition $2k\theta \geq \sigma^2$, see for instance \cite{Jeanblanc}, the origin is inaccessible. These two properties combined with its analytical tractability make the CIR model well-suited for a non-negative interest rate setting.

There is a rich literature on extensions to the classical CIR model in order to obtain more sophisticated models, which could fit the market data better, allowing to price interest rate derivatives more accurately. For example, Chen in \cite{Chen} proposed a three-factor model; Brigo and Mercurio in  \cite{BrigoMercurioLibro} proposed a jump diffusion model (JCIR). In order to include time dependent coefficients in \eqref{eq:CIR}, 
Brigo and Mercurio in \cite{BrigoMercurioPaper} proposed to add a deterministic function into equation \eqref{eq:CIR}. This model, called CIR++, is able to fit the observed term structure of interest rates exactly, while preserving the positivity of the process $r(t)$. Brigo and El-Bachir in \cite{BrigoCDS} generalized the CIR++ model by adding a jump term described by a time-homogeneous Poisson process and Brigo and Mercurio in \cite{BrigoMercurioLibro} studied the CIR2++ model. Another way to generalize the CIR model by including time dependent coefficients in equation \eqref{eq:CIR} was introduced
by Jamshidian in \cite{Jamshidian} and Maghsoodi in \cite{Maghsoodi}, which are known as extended CIR models.

But in the last decade the financial industry encountered a paradigm shift by allowing the possibility of negative interest rates, making the classical CIR model unsuitable.

One way to handle the challenges entailed by negative interest rates is to use Gaussian models with one or more factors, such as the Hull and White model (see \cite{HullWhite}), which also has a very good analytical tractability. A generalization of these models with a good calibration to swaption market prices was found in \cite{DiFrancesco}, while Mercurio and Pallavicini in \cite{MercurioPallavicini} proposed a mixing Gaussian model coupled with parameter uncertainty.

But the glamor of the CIR model is still alive even in the current market environment with negative interest rates. Orlando et al. suggest in several papers (cf. \cite{Orlando2}, \cite{Orlando1} and  \cite{Orlando3}) a new framework, which they call CIR\# model, that fits the term structure of interest rates. Additionally, it preserves the market volatility, as well as the analytical tractability of the original CIR model. Their new methodology consists in partitioning the entire available market data sample, which usually consists of a mixture of probability distributions of the same type.
They use a technique to detect suitable sub-samples with normal or gamma distributions. In a next step, they calibrate the CIR parameters to shifted market interest rates, such that the interest rates are positive, and use a Monte Carlo scheme to simulate the expected value of interest rates.

In this paper, however, we introduce a new methodology for handling the challenges arising from negative interest rates. In our model, the instantaneous spot rate is defined as the difference between two independent classical CIR processes, which allows the preservation of the analytical tractability of the original CIR model without introducing any shift to the market interest rates.


The paper is organized as follows. In \Cref{sec:model} we introduce the model in a general affine model setup and describe our main result \Cref{thm:main}. We will derive the Riccati equations associated with the proposed term structure suitable for the difference of two independent affine processes and solve those explicitly in a CIR framework. 

After that, in \Cref{sec:numerics}, we will conduct some numerical experiments. First, we calibrate our model via \eqref{eq:termstructure} to the market data at \dateA and \dateB in \Cref{sec:calibration}. Subsequently, we simulate the model by using the Euler-Maruyama scheme in \Cref{sec:emc} and study the mean, variance and distribution of the model in \Cref{sec:mean_var}. Then we test how the calibrated model performs when pricing zero-coupon bonds at future times in \Cref{sec:forward_zcprices} and conclude our numerical tests by pricing swaptions in \Cref{sec:pricing_swaps}.
Finally, we summarize the results of the paper in \Cref{sec:conclusion} and discuss possible extensions for future research.

\section{A model for negative interest rates}\label{sec:model}
We will now describe how \Cref{thm:main} can be derived. As aforementioned, we consider all dynamics under the risk-neutral measure $\Q$ and give now a heuristic argument, why it makes sense to choose the term structure in \Cref{thm:main} as in \eqref{eq:termstructure}.

Suppose, that $x(t)$ and $y(t)$ are both independent affine processes. 
Then the the price of a zero-coupon bond for each of them separately (cf. \cite{BrigoMercurioLibro} p.~69) is given by
\begin{align}
	P(t,T)=E^\Q_t\left[e^{-\int_{t}^{T}{z(s)ds}}\right]=A_z(t,T)e^{-B_z(t,T)z(t)},
	\label{eq:PtTz}
\end{align}
where $z\in \left\{x,y\right\}$ and $E^\Q_t$ denotes the conditional expectation with respect to $\mathcal{F}_t$ under the measure $\Q$. Now, consider $r(t)=x(t)-y(t)$, then
we have by linearity and independence
\begin{align*}
 P(t,T)	 =  E^\Q_t\left[e^{-\int_{t}^{T}r(s) ds}\right] 
				 =  E^\Q_t\left[e^{-\int_{t}^{T}(x(s) - y(s)) ds}\right] 
				 =  E^\Q_t\left[e^{-\int_{t}^{T}x(s) ds}\right] 
						E^\Q_t\left[e^{\int_{t}^{T}y(s) ds}\right].
\end{align*}
If we concentrate in \eqref{eq:PtTz} only on the right-hand side, it would make sense for two independent processes $x$ and $y$ that we can apply these formulas with a change of sign in front of $B_y$, leading to
\begin{align*}
	P(t,T) \overset{!}{=}  A_x(t,T)e^{-B_x(t,T)x(t)}A_y(t,T)e^{B_y(t,T)y(t)}.
\end{align*}
In the following Lemma we will make this argument rigorous.
\begin{lemma}\label{lem:Riccati}%
	Let everything be as in \Cref{thm:main} but let $x(t)$ and $y(t)$ follow the general
	affine dynamics described in \eqref{eq:x} and \eqref{eq:y}.
	
	Then, the price of a Zero-coupon bond is given by
	\begin{align}
		P(t,T)=
		E^\Q_t\left[e^{-\int_{t}^{T}r(s) ds}\right]=
		A_x(t,T)e^{-B_x(t,T)x(t)}A_y(t,T)e^{B_y(t,T)y(t)},
		\label{eq:PtTaffine}
	\end{align}
	where $A_z$ and $B_z$, $z\in \left\{x,y\right\}$, are deterministic functions and are a classical solution to the following system of Riccati equations
	\begin{align}
	\left\{
	\begin{aligned}[c]\arraycolsep=0pt
		-1 - B_x(t,T) \lambda_x(t) - \left(\partial_t B_x\right)(t,T) 
				+\frac{1}{2}B_x^2(t,T)\gamma_x(t)&=0,\quad B_x(T,T)=0\\
		 -B_x(t,T)\eta_x(t) + \frac{1}{2} B_x^2(t,T) \delta_x(t) + \partial_t\left(\log A_x\right)(t,T)&=0,\quad A_x(T,T)=1\\
		1 + B_y(t,T) \lambda_y(t) + \left(\partial_t B_y\right)(t,T) 
				+\frac{1}{2}B_y^2(t,T)\gamma_y(t)&=0,\quad B_y(T,T)=0\\
		 B_y(t,T)\eta_y(t) + \frac{1}{2} B_y^2(t,T) \delta_y(t) + \partial_t\left(\log A_y\right)(t,T)&=0,\quad A_y(T,T)=1.
	\end{aligned}
	\right.
	\label{eq:RiccatiEq}
	\end{align}
\end{lemma}
The proof of this Lemma is referred to \Cref{sec:Riccati}. 
The independence of $x$ and $y$ ensures that the Riccati equations for $A_x$ and $B_x$ are decoupled from the ones for $A_y$ and $B_y$, making it possible to use the existing literature on explicit solutions in the context of short rate models to construct easily a solution for our difference process \eqref{eq:r} in the case where $x$ \eqref{eq:x} and $y$ \eqref{eq:y} are CIR processes.
\begin{remark}\label{rem:extensions}%
	One can immediately use \Cref{lem:Riccati} and the ideas in \Cref{sec:Riccati} to construct solutions to other popular one-factor affine short rate models, where an explicit solution is available, e.g. the Vasicek model, provided that $x$ and $y$ are independent.
	
	Introducing dependence between $x$ and $y$ suggests a coupling of $A_x$ and $B_x$ to $A_y$ and $B_y$ and might have an impact on the analytical tractability, but is left for future research.
\end{remark}

\begin{figure}%
\includegraphics[width=\columnwidth]{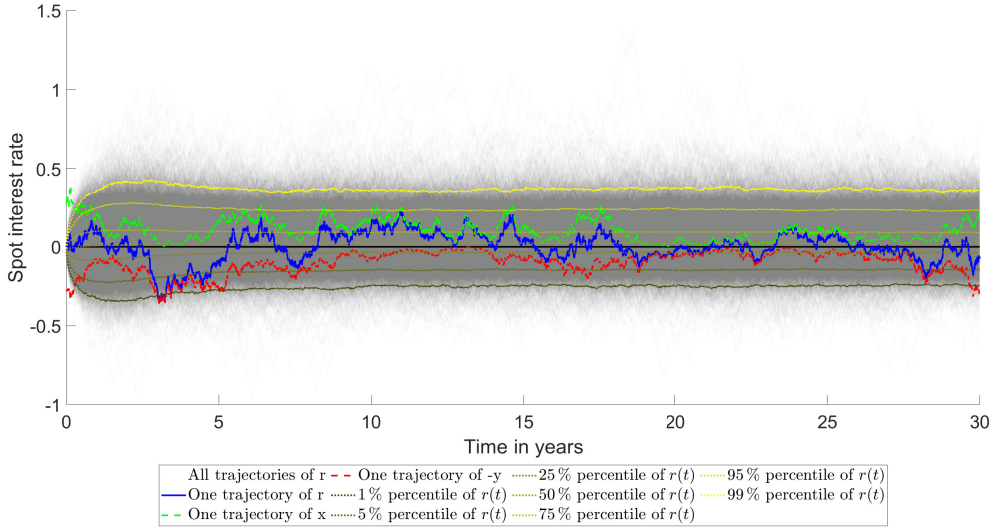}
\caption{An example of a trajectory with negative interest rates $r=x-y$ and its decomposition in $x$ and $-y$, obtained with the market data on \dateA and parameters given in \Cref{tab:risultati_cal}.}%
\label{fig:R_1_A}%
\end{figure}

It is well-known that, under the conditions $2k_z\theta_z \geq \sigma_z^2  
$, $z\in \left\{x,y\right\}$, the processes $x(t)$ and $y(t)$ are strictly positive for every $t\geq 0$ (see for instance \cite{CIR} or \cite{Jeanblanc}). We underline that even if the processes $x(t)$ and $y(t)$ are strictly positive, the instantaneous spot rate $r(t)$ could be negative since it is defined as the difference of $x(t)$ and $y(t)$ for every $t>0$, which is illustrated in \Cref{fig:R_1_A} together with several percentiles of $r(t)$.
\section{Numerical tests}\label{sec:numerics}
We will now perform some numerical experiments in our model.
In \Cref{sec:market_data} we will briefly discuss the market data, which we will use to perform all numerical tests in the subsequent sections. Afterwards, we will describe the calibration procedure of our model to the zero-coupon curves at \dateA and \dateB in \Cref{sec:calibration}. This is followed by a short subsection on simulating the model with the Euler-Maruyama scheme in \Cref{sec:emc} and we investigate the mean, variance and distribution of the short rate model in \Cref{sec:mean_var}.
In \Cref{sec:forward_zcprices} we price zero-coupon bonds at future dates and compare the results to the market prices. Last but not least, \Cref{sec:pricing_swaps} will show results on pricing swaptions in our model.

We used for the calculations \matlab with the \matlabGOtoolbox
running on \OS, on a machine with the following specifications: processor
\CPU and \RAM.
\subsection{Market Data}\label{sec:market_data}
To obtain the market zero-coupon bond term structure, we first build the EUR Euribor-swap curve which is created from the most liquid interest rate instruments available in the market and constructed as follows: We consider deposit rates and Euribor rates with maturity from one day to one year and par-swap rates versus six-month Euribor rates with maturity from two years to thirty years. Then the zero interest curve and the zero-coupon bond curve are calculated using a standard ``bootstrapping'' technique in conjunction with cubic spline interpolation of the continuously compounded rate (cf. \cite{Miron1991} for more details). 

We choose two different dates and we take the data at the end of each business day. In particular, we test our model at \dateA and at \dateB. At the first date, the zero interest rates were negative up to year six, while at the second date the entire zero interest rate structure was negative. In \Cref{tab:first_curve} and in \Cref{tab:second_curve} we report the zero interest rate curve and the zero-coupon bond curve at the two different dates.

Furthermore, in \Cref{sec:pricing_swaps} we need the strikes to compute the model swaption prices and the market prices of the swaptions to compare our model, which are for both dates in the Appendix in 
\Cref{tab:swapPrices_1}, \Cref{tab:swapPrices_2}, \Cref{tab:swapStrike_1} and \Cref{tab:swapStrike_2}, respectively.

All data has been downloaded from Bloomberg and is used in the following subsections for our numerical experiments. We start in the next subsection with calibrating our model to the zero-coupon curve.
\subsection{Calibration}\label{sec:calibration}
In this subsection we will discuss how we calibrate our model to the market zero-coupon curve given in \Cref{tab:first_curve} and \Cref{tab:second_curve} by using the formula derived in \eqref{eq:termstructure}.

Let us denote $\Pi \coloneqq \left[\phi^x_1,\phi^x_2,\phi^x_3,\phi^y_1,\phi^y_2,\phi^y_3,x_0,y_0\right]^T \in \R^8$. We will formulate the calibration procedure as a constraint minimization problem in $\R^8$
for the parameters $\Pi$ with objective function
\begin{align}
	f(\Pi)\coloneqq
	\sum_{i=1}^{n}{\left(\frac{P^M\left(0,T_i\right)}{P\left(\Pi;0,T_i\right)}-1\right)^2},
	\label{eq:objective}
\end{align}
where $n\in\N$ is the number of time points, where market data is available, and $T_i$, $i=1,\dots,n$ are these maturities.
The market zero-coupon curve is denoted by $P^M(0,T_i)$ and $P(\Pi;0,T_i)$ is the price of a zero-coupon bond in our model given by \eqref{eq:termstructure} with parameters $\Pi$.

The objective function describes the relative square difference between the market zero-coupon bond prices and the theoretical prices from the model given by \eqref{eq:termstructure}.

The set of admissible parameters $\mathcal{A}$ will consist of the following constraints arising from the well-definedness of the formulas \eqref{eq:phi}:
\begin{compactenum}
	\item First of all, let us note that there is a one-to-one correspondence between the parameters $\Pi$ and $k_z$, $\sigma_z$ and $\theta_z$ if one is looking for positive real solutions only. We have
	\begin{align}
		\begin{aligned}[c]\arraycolsep=0pt
		k_x &= 2 \phi^x_2 - \phi^x_1,\qquad 
		&&k_y = 2 \phi^y_2 - \phi^y_1,\\
		\sigma_x &= \sqrt{2\left(\phi^x_2\phi^x_1-\left(\phi^x_2\right)^2\right)},\qquad
		&&\sigma_y = \sqrt{-2\left(\phi_2^y\phi_1^y-\left(\phi_2^y\right)^2\right)},\\
		\theta_x &= 
			-\frac{\phi^2_x\phi^3_x(\phi^1_x - \phi^2_x)}{\phi^1_x - 2\phi^2_x},\qquad
		&&\theta_y = 
			\frac{\phi^2_y\phi^3_y(\phi^1_y - \phi^2_y)}{\phi^1_y - 2\phi^2_y}.
		\end{aligned}
		\label{eq:modelParam}
	\end{align}
	\item\label{item:cond1} We require  $\sigma_z \in \R_{\geq 0}$, $z\in \left\{x,y\right\}$. 
	 By rearranging \eqref{eq:modelParam}, these conditions are equivalent to
	$\phi_1^x \geq \phi_2^x$ and $\phi_2^y \geq \phi_1^y$;
	\item A positive mean-reversion speed, i.e. $k_z \geq 0$, is equivalent to 
		$2\phi_2^z \geq \phi_1^z$, $z\in \left\{a,b\right\}$;
	\item The Feller condition $2k_z\theta_z \geq \sigma_z^2 $ is equivalent to 
		$\phi_3^z\geq 1$, $z\in \left\{a,b\right\}$;
	\item A positive mean for each CIR process, i.e. $\theta_z \geq 0$, is 
		by positivity of $\sigma_z^2$ and $k_z$ equivalent to $\phi_3^{z}\geq 0$, which is already satisfied by the Feller condition;
	\item The parameter $\phi_1^z$, assuming that it is real-valued, is positive by definition, meaning that by the positivity of the mean reversion speed, $\phi_2^z$ will be as well. Therefore, all $\phi$ are positive;
	\item As both CIR processes $x_t$ and $y_t$, individually, are positive processes, we additionally require $x_0\geq 0$ and $y_0\geq 0$.
\end{compactenum}
The advantage of using the parameters $\Pi$ instead of $k_z$, $\sigma_z$ and $\theta_z$
is that we can rewrite these conditions as a system of linear inequality constraints in matrix notation $A\cdot\Pi\leq 0$, where
\begin{align*}
	A\coloneqq 
	\left[\begin{array}[c]{*{8}{c}}
    -1  &  1  &  0  &  0  &  0  &  0  &  0  &  0\\
     0  &  0  &  0  &  1  & -1  &  0  &  0  &  0\\
     1  & -2  &  0  &  0  &  0  &  0  &  0  &  0\\
     0  &  0  &  0  &  1  & -2  &  0  &  0  &  0\\
	\end{array}\right]
\end{align*}
with boundary conditions $\Pi_i\geq 0$, $i=1,\dots,8$, and $\Pi_3=\phi_3^x\geq 1$, as well as $\Pi_6=\phi_3^y\geq 1$.

In total, the set of admissible parameters is given by 
\begin{align}
	\mathcal{A}\coloneqq 
	\left\{
		\Pi \in \R^8_{\geq 0}, \Pi_3, \Pi_6 \geq 1 : A \cdot \Pi \leq 0
	\right\}.
	\label{eq:constraints}
\end{align} 

Finally, a solution $\Pi^*$ to the calibration problem is a minimizer of
\begin{align}
	\min_{\Pi \in \mathcal{A}} f\left(\Pi\right).
	\label{eq:min}
\end{align}

To solve \eqref{eq:min} numerically, we want to use \verb+Matlab+'s function \fmincon in
the \matlabGOtoolbox. In order to use this function, we need an initial guess of the parameter $\Pi$ and the computational time will depend on that choice. 
In \Cref{tab:ctimeTests} we present a few choices for initial guesses of $\Pi$. The first row for each date \dateA or \dateB refers to \verb+Matlab+'s function \ga in the \matlabGOtoolbox, which uses a generic global optimization algorithm to find 
a solution of \eqref{eq:min} without starting from an initial guess, which takes a long time to compute, roughly 35 to 43 seconds.
In the following three rows are three manual initial guesses. We can see that the first two choices work for both dates exceptionally fast (0.3 seconds) and the accuracy is almost identical to all other choices, making this model a good choice if live calibration to the data is needed, which we also use in the following numerical experiments. In the last two rows we used random starting parameters to demonstrate that the error remains stable but the computational time varies.

For the algorithms used by \verb+Matlab+ we refer the reader to \cite{Gilli}, in the context of financial mathematics.

\begin{table}%
\caption{Calibration times and corresponding mean relative errors (MRE) for different initial parameters at \dateA and \dateB.}
\centering
\begin{tabular}{|ccc|}
\hline
\ctimeTestsAHeader
\hline
\multicolumn{3}{|c|}{\bfseries Calibration at \dateA}\\
\ctimeTestsABody\\
\hline
\multicolumn{3}{|c|}{\bfseries Calibration at \dateB}\\
\ctimeTestsBBody\\
\hline
\end{tabular}
\label{tab:ctimeTests}
\end{table}

The results of the aforementioned calibration procedure are displayed in \Cref{tab:risultati_cal} for both dates \dateA and \dateB. On the left-hand side, one can see the parameters $\Pi^*$ and
on the right-hand side the corresponding model parameters derived from $\Pi^*$.
At both dates we obtain good results in fitting the market term structure. The mean relative error (MRE), i.e.
$
	\frac{1}{n}\sum_{i=1}^{n}{\abs{\frac{P^M\left(0,T_i\right)}{P\left(\Pi^*;0,T_i\right)}-1}},
$ 
over the entire term structure is $\MREA$ at the first date and $\MREB$ at the second date.

\begin{table}
	\centering
	\caption{Calibration parameters $\Pi^*$, model parameters and mean relative errors (MRE) at \dateA and \dateB, obtained with the market data given in \Cref{tab:first_curve} and \Cref{tab:second_curve}.}
  \begin{tabular}{@{}|c|*{2}{c}|@{}}
  \hline 
	Parameter  	 & \dateA & \dateB\\
	\hline
	$\phi_1^x$ 	 & $\phiOneXA$ & $\phiOneXB$\\
	$\phi_2^x$ 	 & $\phiTwoXA$ & $\phiTwoXB$\\
	$\phi_3^x$ 	 & $\phiThreeXA$ & $\phiThreeXB$\\
	$x_0$ 		 	 & $\xZeroA$ & $\xZeroB$\\
	$\phi_1^y$ 	 & $\phiOneYA$ & $\phiOneYB$\\
	$\phi_2^y$ 	 & $\phiTwoYA$ & $\phiTwoYB$\\
	$\phi_3^y$ 	 & $\phiThreeYA$ & $\phiThreeYB$\\
	$y_0$ 		 	 & $\yZeroA$ & $\yZeroB$\\
	\hline
	$f\left(\Pi^*\right)$ & $\errFminA$ & $\errFminB$\\
	\textbf{MRE} & $\MREA$ & $\MREB$\\
	\hline
  \end{tabular}%
	\begin{tabular}{@{}|c|*{2}{c}|@{}}
  \hline 
	Parameter  	 & \dateA & \dateB\\
	\hline
	$k_x$ 	 		 & $\kXA$ & $\kXB$\\
	$\sigma_x$ 	 & $\sigmaXA$ & $\sigmaXB$\\
	$\theta_x$ 	 & $\thetaXA$ & $\thetaXB$\\
							 & 					   & 					\\
	$k_y$ 			 & $\kYA$ & $\kYB$\\
	$\sigma_y$ 	 & $\sigmaYA$ & $\sigmaYB$\\
	$\theta_y$ 	 & $\thetaYA$ & $\thetaYB$\\
							 & 					  & 					 \\
	\hline
	  &  & \\
	 &  & \\
	\hline
  \end{tabular}
  \label{tab:risultati_cal}
\end{table}
\begin{figure}
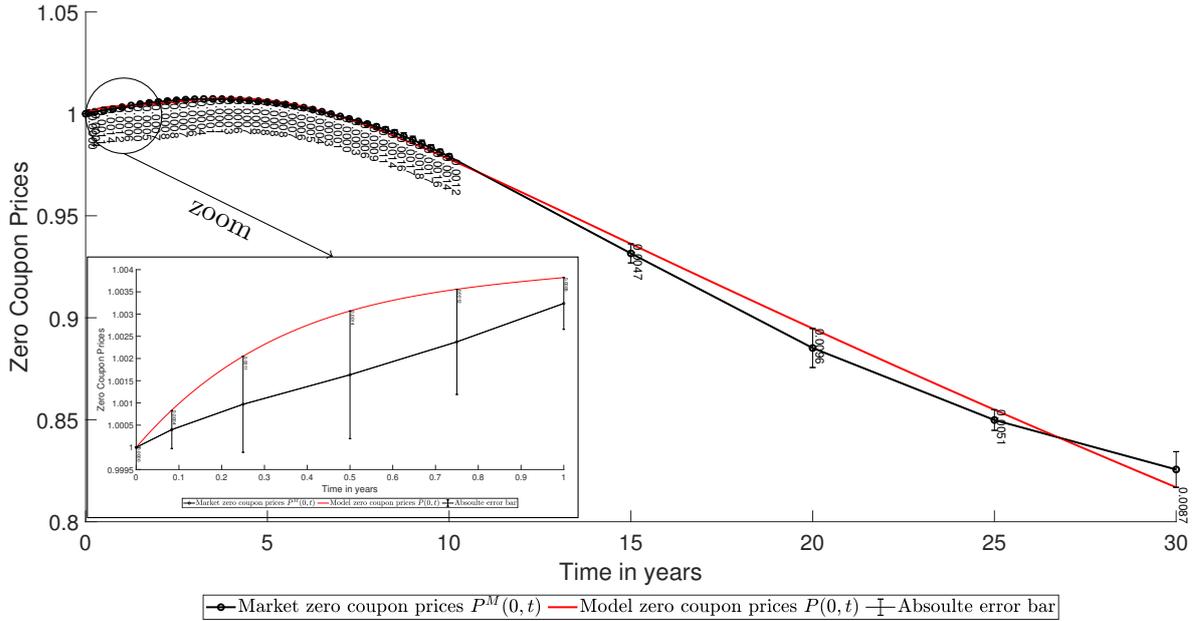
%
\begin{tikzpicture}
\node[] (figZCOneA) at (0,0) {\includegraphics[width=\columnwidth]{\figZCOneA}};
\node[draw=black] (figZCTwoA) at (-3.5,-1) {\includegraphics[width=.4\columnwidth]{\figZCTwoA}};
\draw[black] (-6.25,2.6) circle (.5cm) node (spyZCOneA) {};
\draw[thin,->] ($(spyZCOneA)+(0,-.5cm)$) -- (figZCTwoA.north) node[midway,sloped, anchor=center,below] {zoom};
\end{tikzpicture}
\caption{A comparison of the market zero-coupon prices (\Cref{tab:first_curve}) to the model zero-coupon prices with absolute errors at \dateA with parameters given in \Cref{tab:risultati_cal}.}%
\label{fig:ZC_1_A}%
\end{figure}
\subsection{Euler-Monte-Carlo simulation}\label{sec:emc} 
In order to forecast the future expected interest rate, we use the Euler-Maruyama scheme to simulate the instantaneous spot rate $r$ \eqref{eq:r}. We refer to \cite{Dereich} and the references therein for a list of different Euler-type methods to simulate a CIR process. In our experiments, we simulate the processes $x(t)$ and $y(t)$ by the truncated Euler scheme defined as follows:

First of all, we fix a homogeneous time grid $0=t_0\leq t_1 \leq \cdots \leq t_N=T$ 
for the interval $[0,T]$ with $N+1$ time points and mesh $\Delta t_i \coloneqq t_{i+1}-t_{i} \equiv \Delta \coloneqq \frac{T}{N}$ for all $i=0,\dots,N-1$. Secondly, we simulate the two independent Brownian motions $W_z$, $z\in \left\{x,y\right\}$, and define their time increment as $\Delta W_{z}(t_i)\coloneqq W_{z}(t_{i+1})-W_{z}(t_{i})$.
In total, we compute $r(t_{i+1})\coloneqq x(t_{i+1})-y(t_{i+1})$ for $i=0,\dots,N-1$, where
\begin{align}\label{eq:euler}
\begin{aligned}[c]\arraycolsep=0pt
 	x({t_{i+1}}) &= x({t_i}) + k_x(\theta_x - x({t_i}))\Delta t_i + \sigma_x\sqrt{\max(x({t_i}),0)}\Delta W_x({t_i}) \\
 	y({t_{i+1}}) &= y({t_i}) + k_y(\theta_y - y({t_i}))\Delta t_i + \sigma_y\sqrt{\max(y({t_i}),0)}\Delta W_y({t_i}).
\end{aligned}
\end{align}
We choose the $\max$ inside the square-root to ensure that the square-root remains real, because due to discretization effects the positivity of $x({t_i})$ and $y({t_i})$ might be violated.

In the following experiments we choose $\Delta=\timeMeshA$ and use $M=\simulationsA$ samples for each of the Brownian motions. In \Cref{fig:DF_1_A} we show the mean and 
$99.9\,\%$ confidence interval (under the assumption of the central limit theorem) of the model discount factors compared to the market discount factors at \dateA.
One can see that the mean does not differ from the market discount factors very much till 5 years with an error of magnitude $0.005$ and increases slightly to a magnitude of $0.05$
afterwards till 30 years.
\begin{figure}%
\begin{tikzpicture}
\node[] (figDFOneA) at (0,0) {\includegraphics[width=\columnwidth]{\figDFOneA}};
\node[draw=black] (figDFTwoA) at (-3.5,-.65) {\includegraphics[width=.4\columnwidth]{\figDFTwoA}};
\draw[black] (-6.25,3) circle (.5cm) node (spyDFOneA) {};
\draw[thin,->] ($(spyDFOneA)+(0,-.5cm)$) -- (figDFTwoA.north) node[midway,sloped, anchor=center,below] {zoom};
\end{tikzpicture}
\caption{A comparison of the market discount factors (\Cref{tab:first_curve}) to the mean of the model discount factors with absolute errors at \dateA with parameters given in \Cref{tab:risultati_cal} by using \eqref{eq:euler} with $\Delta=\timeMeshA$ and $M=\simulationsA$.}%
\label{fig:DF_1_A}%
\end{figure}

A more detailed comparison of the mean absolute errors, i.e. the absolute value of the difference of the mean over all simulations of our model to the market data, at each maturity can be found in the appendix in \Cref{tab:dfMeanErrAB}.
  
\subsection{Mean and variance}\label{sec:mean_var}
The $\mathcal{F}_s$-conditional mean and variance of the CIR process are well-known (cf. \cite{BrigoMercurioLibro} p.~66 equation (3.23)) and are given by
\begin{align*}
	E^\Q_s\left[
			z(t)
	\right]&=
	z(s)e^{-k_z(t-s)}+
	\theta_z 
		\left(
			1-e^{-k_z(t-s)}
		\right)
	\\
	\mathrm{Var}^\Q_s\left[
			z(t)
	\right]&=
	z(s)\frac{\sigma_z^2}{k_z}
		\left(
			e^{-k_z(t-s)}
			-e^{-2k_z(t-s)}
		\right)+
	\theta_z \frac{\sigma_z^2}{2k_z}
		\left(
			1 - e^{-k_z(t-s)}
		\right)^2,
\end{align*}
where $z\in \left\{x,y\right\}$. In the case of the difference of two CIR processes we have
\begin{align}
	E^\Q_s\left[
			r(t)
	\right]&=
		E^\Q_s\left[
			x(t)-y(t)
	\right]
	=E^\Q_s\left[
			x(t)
	\right]-
	E^\Q_s\left[
			y(t)
	\right]
	\label{eq:meanCIRminus}
\end{align}
and by independence
\begin{align}
	\mathrm{Var}^\Q_s\left[
			r(t)
	\right]&=
	\mathrm{Var}^\Q_s\left[
			x(t)
	\right]+
	\mathrm{Var}^\Q_s\left[
			y(t)
	\right].
	\label{eq:varCIRminus}
\end{align}
\begin{figure}
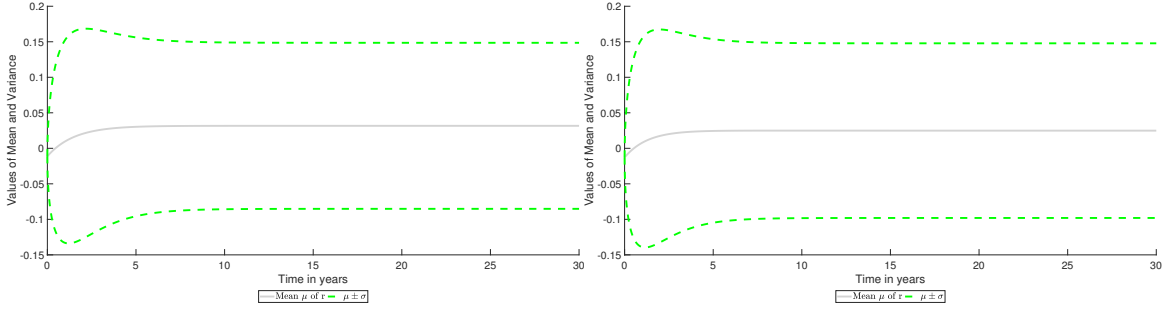
%
\includegraphics[width=.49\columnwidth]{\figMeanVarOneA}%
\includegraphics[width=.49\columnwidth]{\figMeanVarOneB}%
\caption{Mean and standard deviation of $r(t)$ using the calibrated parameters in \Cref{tab:risultati_cal}, obtained with \eqref{eq:meanCIRminus} and \eqref{eq:varCIRminus}. The left picture shows the results at \dateA and the right at \dateB.}%
\label{fig:distribution}%
\end{figure}
In \Cref{fig:distribution} we show for each date the histogram of the short rate distribution after $30$ years. To describe the distribution of $r(t)$ after $30$ years better, we also compare it to the density of a normal random variable with the same mean and variance. As one expects, the distribution of $r$ shows a slight skewness and fatter tail with respect to the normal distribution. 
\begin{figure}%
\includegraphics[width=.49\columnwidth]{\figMeanVarThreeA}%
\includegraphics[width=.49\columnwidth]{\figMeanVarThreeB}%
\caption{Distribution of the simulated short rate $r(t)$ compared to the normal distribution at $t=30$ using the calibrated parameters in \Cref{tab:risultati_cal}, $\Delta=\timeMeshA$ and $M=\simulationsA$. The left picture shows the results at \dateA and the right at \dateB.}%
\label{fig:distribution}%
\end{figure}

\subsection{$t$-Forward zero-coupon prices}\label{sec:forward_zcprices}
In this subsection we examine, if our model is able to replicate the interest rate term structure not only at the start date but also in the future. Therefore, we compare the market and model prices of zero-coupon bonds at future times $t$ by using the continuously-compounded spot interest rates (cf. \cite{BrigoMercurioLibro} Definition 1.2.3.) to obtain the market prices.

To be more precise, let us briefly recall the definition of the continuously-compounded spot interest rates
\begin{align}
	R(t,T)\coloneqq -\frac{\log P(t,T)}{T-t},
	\label{eq:ccr}
\end{align}
where we use the day-count convention $T-t$ in years.

To compute this rate for the market at a fixed future date $t>0$ with maturity $T>t$ we use
\begin{equation}\label{eq:forward}
  R^M(t,T) \coloneqq \frac{T R^M(0,T) - t R^M(0,t)}{T-t}
\end{equation} 
where $R^M(0,t)$ and $R^M(0,T)$ are the market zero rates with maturity $t$ and $T$ respectively. 

By rearranging \eqref{eq:ccr} we derive the market zero-coupon price at a future date $t$
with maturity $T$ as 
\begin{align}
	P^M(t,T)= \exp\left(- R^M(t,T) (T-t)\right).
	\label{eq:marketforwardzcprice}
\end{align}

Now, we can compare \eqref{eq:marketforwardzcprice} to the mean over all simulated trajectories of the model $t$-forward zero-coupon price derived from \eqref{eq:termstructure}.

In \Cref{fig:F_1}, \Cref{fig:F_2} and \Cref{fig:F_3} we show a comparison of \eqref{eq:marketforwardzcprice} to \eqref{eq:termstructure} at \dateA and \dateB with parameters 
$\Pi^*$ from \Cref{tab:risultati_cal} for future times $t=1,3,5$, respectively.

The behavior shown by the model $t$-forward prices coincides with typical one-factor short interest rate models. 
At short future dates, one year for example, the model is able to reproduce the market forward zero-coupon price for \dateA, i.e. the error is of magnitude $0.01$, whereas at long future dates, e.g. \Cref{fig:F_3}, the model prices are deviating further from the market prices, i.e. the error is of magnitude $0.03$. For \dateB we can see that the errors are not very large but the model seems to have difficulties to match the shape of the predicted market prices.

\begin{figure}%
\includegraphics[width=.49\columnwidth]{\figFOneA}%
\includegraphics[width=.49\columnwidth]{\figFOneB}%
\caption{Mean and $99\,\%$ confidence interval of the model $t$-forward zero-coupon prices compared to the market $t$-forward zero-coupon prices at $t=1$ using the calibrated parameters in \Cref{tab:risultati_cal}, $\Delta=\timeMeshA$ and $M=\simulationsA$. The left picture shows the results at \dateA and the right at \dateB.}%
\label{fig:F_1}%
\end{figure}
\begin{figure}%
\includegraphics[width=.49\columnwidth]{\figFTwoA}%
\includegraphics[width=.49\columnwidth]{\figFTwoB}%
\caption{Mean and $99\,\%$ confidence interval of the model $t$-forward zero-coupon prices compared to the market $t$-forward zero-coupon prices at $t=3$ using the calibrated parameters in \Cref{tab:risultati_cal}, $\Delta=\timeMeshA$ and $M=\simulationsA$. The left picture shows the results at \dateA and the right at \dateB.}%
\label{fig:F_2}%
\end{figure}
\begin{figure}%
\includegraphics[width=.49\columnwidth]{\figFThreeA}%
\includegraphics[width=.49\columnwidth]{\figFThreeB}%
\caption{Mean and $99\,\%$ confidence interval of the model $t$-forward zero-coupon prices compared to the market $t$-forward zero-coupon prices at $t=5$ using the calibrated parameters in \Cref{tab:risultati_cal}, $\Delta=\timeMeshA$ and $M=\simulationsA$. The left picture shows the results at \dateA and the right at \dateB.}%
\label{fig:F_3}%
\end{figure}
\subsection{Pricing swaptions}\label{sec:pricing_swaps}
In this subsection we test if our model is market consistent, in the sense whether the model is able to reproduce market swaption prices or not.

We compare market swaption prices to model swaption prices with different tenors ($1,2,5,7,10$ years) and maturities ($1,2,5,7,10,15,20$ years). 
The market swaption prices are computed by Bachelier's formula from normal volatilities quoted in the market whereas the model swaption prices are from the simulated future zero-coupon prices in \eqref{eq:termstructure}. 
The difference between market price to model prices for \dateA and \dateB are reported in \Cref{tab:swaption1} and \Cref{tab:swaption2}, respectively. We notice that, similar to one-factor short interest rate model, our model fails to capture the full swaption volatility surface. This result is not surprising, since the model uses essentially a single volatility factor due to the fact that the model parameters are constant and the Brownian motion are independent. A way to make the model able to match the entire volatility surface is the inclusion of time-dependent parameters to fit exactly the market term structure or, equivalently, adding a deterministic function of time to the short rate process $r$, which is left for future research. 
\begin{table}
	\centering
	\caption{\dateA: difference between swaption model price and swaption market price}
	\errorSwaptionA
  \label{tab:swaption1}
\end{table}
\begin{table}
	\centering
	\caption{\dateB: difference between swaption model price and swaption market price}
	\errorSwaptionB
  \label{tab:swaption2}
\end{table}
\section{Conclusion and future research}\label{sec:conclusion}
In this paper, we propose a new model to handle the challenges arising from negative interest rates, while preserving the analytical tractability of the original CIR model without introducing any shift to the market interest rates. The strength of our model is that it is very simple, fast to calibrate and fits the present market term structure very well for an essentially one-factor short rate model. 

Let us briefly summarize our discoveries of the numerical section.
We show that the distribution of the short rate after $30$ years has similar features compared to the original CIR model in terms of skewness and fat tail.
At \dateA we show that the model is quite capable of pricing zero-bonds at future times, while for \dateB the error is not too large but an improvement of the current model would be desirous to allow a better fit to the shape of the market prices.
Similarly, we show that we require an extension of the model to price swaptions more accurately.  

That being said, the authors would like to stress that this is a first step in this methodology of using two CIR processes as a difference to model negative interest rates. Naturally, its extensions, such as considering time-dependent coefficients to fit the market term structure perfectly or, equivalently, adding a deterministic shift extension in the sense of \cite{BrigoMercurioLibro}, will be a next step and is left for future research.

\appendix
\section{Derivation of the Riccati equations and coefficients}\label{sec:Riccati}
Let everything be as in \Cref{lem:Riccati}. In particular, let
\begin{align*}
	P(t,T) \overset{!}{=}
	A_x(t,T) \exp\left(-B_x(t,T)x(t)\right) 
	A_y(t,T) \exp\left(B_y(t,T)y(t)\right).
\end{align*}

To derive the Riccati equations \eqref{eq:RiccatiEq} we use the fact, that we are modelling under the martingale measure $\Q$, therefore the discounted price process 
$\exp\left(-\int_{0}^{t}{r(s)ds}\right) P(t,T)$ needs to be a martingale. 

By independence of $x$ and $y$, as well as Itô's formula we derive after some algebra
\begin{align*}
	\hspace{1em}&\hspace{-1em}
	d \left[
		\exp\left(-\int_{0}^{t}{r(s)ds}\right)
		A_x(t,T) \exp\left(-B_x(t,T)x(t)\right) 
		A_y(t,T) \exp\left(B_y(t,T)y(t)\right)
	\right]
	\\&=
	\begin{aligned}[t]\arraycolsep=0pt
		\hspace{1em}&\hspace{-1em}
		A_x(t,T) \exp\left(-\int_{0}^{t}{x(s)ds}-B_x(t,T)x(t)\right) 
		\Biggl[
				\exp\left(\int_{0}^{t}{y(s)ds} +B_y(t,T)y(t)\right)
				\biggl[\\&
					A_y(t,T) 
						\biggl[
							y(t)dt + B_y(t,T) dy(t) + y(t) \left(\partial_t B_y\right)(t,T) dt 
							+
							\frac{1}{2} B_y^2(t,T) d\left\langle y\right\rangle_t
						\biggr]
					\\&+
					\left(\partial_t A_y\right)(t,T) dt
				\biggr]
		\Biggr]
	\end{aligned}
	\\&\quad+
	\begin{aligned}[t]\arraycolsep=0pt
		\hspace{1em}&\hspace{-1em}
		A_y(t,T) \exp\left(\int_{0}^{t}{y(s)ds} +B_y(t,T)y(t)\right)
		\Biggl[
				\exp\left(-\int_{0}^{t}{x(s)ds} +B_x(t,T)x(t)\right)
				\biggl[\\&
					A_x(t,T) 
						\biggl[
							-x(t)dt - B_x(t,T) dx(t) - x(t) \left(\partial_t B_x\right)(t,T) dt 
							+
							\frac{1}{2} B_x^2(t,T) d\left\langle x\right\rangle_t
						\biggr]
					\\&+
					\left(\partial_t A_x\right)(t,T) dt
				\biggr]
		\Biggr].
	\end{aligned}
\end{align*}
Now, in order to be a martingale the parts of bounded variation have to vanish, which leads us after rearranging the terms to
\begin{align*}
	\hspace{1em}&\hspace{-1em}
	0 \overset{!}{=}
	y(t)
	\begin{aligned}[t]\arraycolsep=0pt
		\Biggl[
			A_x(t,T)A_y(t,T)
			\biggl[&
				1 + B_y(t,T) \lambda_y(t) + \left(\partial_t B_y\right)(t,T) 
				+\frac{1}{2}B_y^2(t,T)\gamma_y(t)
			\biggr]
		\Biggr]
	\end{aligned}\\&\quad+
	x(t)
	\begin{aligned}[t]\arraycolsep=0pt
		\Biggl[
			A_y(t,T)A_x(t,T)
			\biggl[&
				-1 - B_x(t,T) \lambda_x(t) - \left(\partial_t B_x\right)(t,T) 
				+\frac{1}{2}B_x^2(t,T)\gamma_x(t)
			\biggr]
		\Biggr]
	\end{aligned}\\&\quad+
	\begin{aligned}[t]\arraycolsep=0pt
		\hspace{1em}&\hspace{-1em}
		A_x(t,T)A_y(t,T)
		\Biggl[
			B_y(t,T)\eta_y(t) + \frac{1}{2} B_y^2(t,T) \delta_y(t) -
			B_x(t,T)\eta_x(t) + \frac{1}{2} B_x^2(t,T) \delta_x(t)
		\Biggr]\\&+
		A_x(t,T) \left(\partial_t A_y\right)(t,T)+
		A_y(t,T) \left(\partial_t A_x\right)(t,T).
	\end{aligned}
\end{align*}
Thus, we derive the following Riccati System
		\begin{align*}
			1 + B_y(t,T) \lambda_y(t) + \left(\partial_t B_y\right)(t,T) 
				+\frac{1}{2}B_y^2(t,T)\gamma_y(t)&=0, \quad &&B_y(T,T)=0,\\
			-1 - B_x(t,T) \lambda_x(t) - \left(\partial_t B_x\right)(t,T) 
				+\frac{1}{2}B_x^2(t,T)\gamma_x(t)&=0, \quad &&B_x(T,T)=0,\\
		\begin{multlined}[b]\arraycolsep=0pt
				B_y(t,T)\eta_y(t) + \frac{1}{2} B_y^2(t,T) \delta_y(t) -
				B_x(t,T)\eta_x(t) + \frac{1}{2} B_x^2(t,T) \delta_x(t) \\+
			\partial_t\left(\log A_y\right)(t,T)+
			\partial_t\left(\log A_x\right)(t,T)
		\end{multlined}&=0.
		\end{align*}
		A solution to the last equation can be found by further assuming that 
		the individual $x$ and $y$ 
		parts will be zero, leading to two separate equations
		\begin{align*}
			B_y(t,T)\eta_y(t) + \frac{1}{2} B_y^2(t,T) \delta_y(t) + \partial_t\left(\log A_y\right)(t,T) &= 0, \quad A_y(T,T)=1,\\
			-B_x(t,T)\eta_x(t) + \frac{1}{2} B_x^2(t,T) \delta_x(t) + \partial_t\left(\log A_x\right)(t,T) &= 0, \quad A_x(T,T)=1.
		\end{align*}
		
We will now turn to the special case of the CIR processes \eqref{eq:CIR}. 
We see immediately that the equations for $x$ are in the usual form and 
defining $\lambda_x(t)\equiv -k_x, \eta_x(t)\equiv k_x\theta_x, \gamma_x(t)\equiv\sigma_x^2, \delta_x(t)\equiv 0$ yields the explicit solution from the literature (cf. \cite{BrigoMercurioLibro} p.~66 equation (3.25)).

Concerning the $y$ terms, we make the following educated guess and verify, that it solves the equation:
%
\begin{align*}
	A(t,T) &= 
	\left(
		\frac{
			2h \exp\left((k+h)(T-t)/2\right)
		}{
			2h + (k+h)\left(\exp\left((T-t)h\right)-1\right)
		},
	\right)^\frac{2k\theta}{\sigma^2}\\
	B(t,T) &=
	\frac{
		2\left(\exp\left((T-t)h\right)-1\right)
	}{
		2h + (k+h)\left(\exp\left((T-t)h\right)-1\right)
	}, \\
	h &= \sqrt{k^2-2\sigma^2},
\end{align*}
where we will drop the index for indicating that we are considering the $y$ coefficients for readability and assume that $k^2 \geq 2 \sigma^2$.
\paragraph*{Verification for $B$}
We will first check the formula for the Riccati equation in $B$:

We will now simplify the nominator and the denominator of 
$\partial_t B + \frac{1}{2} \sigma^2 B^2 - kB$, which is given by
\begin{align*}
	\hspace{1em}&\hspace{-1em}
	\partial_t B + \frac{1}{2} \sigma^2 B^2 - kB=\\&
	\frac{\sigma ^2\,{\left(2\,{\mathrm{e}}^{\sqrt{k^2-2\,\sigma ^2}\,\left(T-t\right)}-2\right)}^2}{2\,{\left(\left({\mathrm{e}}^{\sqrt{k^2-2\,\sigma ^2}\,\left(T-t\right)}-1\right)\,\left(k+\sqrt{k^2-2\,\sigma ^2}\right)+2\,\sqrt{k^2-2\,\sigma ^2}\right)}^2}
	\\&-\frac{2\,{\mathrm{e}}^{\sqrt{k^2-2\,\sigma ^2}\,\left(T-t\right)}\,\sqrt{k^2-2\,\sigma ^2}}{\left({\mathrm{e}}^{\sqrt{k^2-2\,\sigma ^2}\,\left(T-t\right)}-1\right)\,\left(k+\sqrt{k^2-2\,\sigma ^2}\right)+2\,\sqrt{k^2-2\,\sigma ^2}}
	\\&-\frac{k\,\left(2\,{\mathrm{e}}^{\sqrt{k^2-2\,\sigma ^2}\,\left(T-t\right)}-2\right)}{\left({\mathrm{e}}^{\sqrt{k^2-2\,\sigma ^2}\,\left(T-t\right)}-1\right)\,\left(k+\sqrt{k^2-2\,\sigma ^2}\right)+2\,\sqrt{k^2-2\,\sigma ^2}}
	\\&+\frac{{\mathrm{e}}^{\sqrt{k^2-2\,\sigma ^2}\,\left(T-t\right)}\,\sqrt{k^2-2\,\sigma ^2}\,\left(k+\sqrt{k^2-2\,\sigma ^2}\right)\,\left(2\,{\mathrm{e}}^{\sqrt{k^2-2\,\sigma ^2}\,\left(T-t\right)}-2\right)}{{\left(\left({\mathrm{e}}^{\sqrt{k^2-2\,\sigma ^2}\,\left(T-t\right)}-1\right)\,\left(k+\sqrt{k^2-2\,\sigma ^2}\right)+2\,\sqrt{k^2-2\,\sigma ^2}\right)}^2}.
\end{align*}

After bringing the terms to the common denominator, we consider now the nominator of this transformation
\begin{align*}
	\hspace{1em}&\hspace{-1em}
	\begin{aligned}[t]\arraycolsep=0pt
		\hspace{1em}&\hspace{-1em}
		\frac{1}{2}\sigma^2 \left(
			2 \exp\left(\sqrt{k^2 - 2 \sigma^2}\tau\right)
			- 2
		\right)^2 
		\\&-
		\left(
			2\sqrt{k^2 - 2 \sigma^2} 
				\exp\left(
					\sqrt{k^2 - 2 \sigma^2}\tau
				\right)+
			k\left(
				2 \exp\left(
					\sqrt{k^2 - 2 \sigma^2} \tau
				\right)
				- 2
			\right)
		\right)
		\\&\qquad
		\left(
			\left(
				\exp\left(
					\sqrt{k^2 - 2 \sigma^2}\tau 
				\right)
				-1
			\right)
			\left(
				k+
				\sqrt{k^2 - 2\sigma^2}
			\right)
			+ 2 \sqrt{k^2 - 2 \sigma^2}
		\right)
		\\&+
			\exp\left(
				\sqrt{k^2 - 2 \sigma^2} \tau
			\right)
		\sqrt{k^2 - 2 \sigma^2}
		\left(
			k+\sqrt{k^2-2\sigma^2}
		\right)
		\left(
			2\exp\left(
				\sqrt{k^2 - 2 \sigma^2}\tau
			\right)
			-2
		\right)
	\end{aligned}\\&=
	2\,h\,k-2\,k^2-2\,k^2\,{\mathrm{e}}^{2\,h\,\tau }+2\,\sigma ^2+4\,\sigma ^2\,{\mathrm{e}}^{h\,\tau }+2\,\sigma ^2\,{\mathrm{e}}^{2\,h\,\tau }-2\,h\,k\,{\mathrm{e}}^{2\,h\,\tau },
\end{align*}
where we substituted $\tau \coloneqq T-t$ and $h\coloneqq \sqrt{k^2 - 2 \sigma^2}$.
The denominator can be simplified in the same way, leading to
\begin{align*}
	2\,k^2-2\,h\,k+2\,k^2\,{\mathrm{e}}^{2\,h\,\tau }-2\,\sigma ^2-4\,\sigma ^2\,{\mathrm{e}}^{h\,\tau }-2\,\sigma ^2\,{\mathrm{e}}^{2\,h\,\tau }+2\,h\,k\,{\mathrm{e}}^{2\,h\,\tau }.
\end{align*}

In total, we see that the denominator differs only by a sign, hence $\partial_t B + \frac{1}{2} \sigma^2 B^2 - kB = -1$, which yields the claim.

\paragraph*{Verification for $A$}
The formula can be derived by just integrating and taking the exponential.

\begin{align*}
	\left(
	\frac{
		2h {\mathrm{e}}^{\frac{1}{2}\tau\left(k+h\right)}
		}{
		\left({\mathrm{e}}^{\tau\,h}-1\right)\left(h+k\right)+2h
		}
	\right)^{\frac{2\,k\,\theta }{\sigma ^2}},
\end{align*}
where $h\coloneqq \sqrt{k^2-2\,\sigma ^2}$ and $\tau \coloneqq T-t$.

Let us just take the logarithm and derivative to verify this formula:
\begin{align*}
	\log\left(A(\tau)\right)&=
	\frac{\tau \,\left(2\,k^2\,\theta +2\,k\,\theta \,h\right)}{2\,\sigma ^2}
	-
	\frac{2\,k\,\theta \,\ln\left({\mathrm{e}}^{\tau \,h}\,h-k+h+k\,{\mathrm{e}}^{\tau \,h}\right)}{\sigma ^2}
	+
	\frac{2\,k\,\theta \,\ln\left(2\,\sqrt{k^2-2\,\sigma ^2}\right)}{\sigma ^2}.
\end{align*}
Now, taking the derivative yields
\begin{align*}
	\partial_\tau \left(\log \left(A\left(\tau\right)\right)\right)&=
	\frac{2\,k\,\theta \,\left({\mathrm{e}}^{\tau \,h}-1\right)}{{\mathrm{e}}^{\tau \,h}\,h-k+h+k\,{\mathrm{e}}^{\tau \,h}}=
	k\theta
	\frac{
		2\left({\mathrm{e}}^{\tau \,h}-1\right)
	}{
		\left({\mathrm{e}}^{\tau \,h}-1\right)\left(h+k\right)+2h
	}.
\end{align*}
After undoing the substitution for $\tau$ this is equal to $k\theta B$, which yields the claim.
\section{Instantaneous forward rate}\label{sec:instantForwardRate}
The definition of the instantaneous forward rate (cf. \cite{BrigoMercurioLibro} p.~13 equation (1.23)) is given by
\begin{align*}
	f(t,T)\coloneqq
	-\partial_T\log\left(P\left(t,T\right)\right).
\end{align*}
By \eqref{eq:termstructure} we therefore have
\begin{align*}
	f(t,T)&=
	-\partial_T\left(\log\left(A_x(t,T)e^{-B_x(t,T) x(t)}A_y(t,T)e^{B_y(t,T)y(t)}\right)\right)\\&=
	-\partial_T
		\left(
		\log\left(A_x(t,T)\right)-B_x(t,T) x(t)
		\right)
	-\partial_T
		\left(
		\log\left(A_y(t,T)\right)+B_y(t,T)y(t)
		\right)\\&=
		-\frac{\partial_T\left(A_x(t,T)\right)}{A_x(t,T)}
		+\partial_T\left(B_x(t,T)\right) x(t)
		-\frac{\partial_T\left(A_y(t,T)\right)}{A_y(t,T)}
		-\partial_T\left(B_y(t,T)\right) y(t).
\end{align*}
Let $z\in \left\{x,y\right\}$ and consider the case of the CIR model \eqref{eq:CIR}. Then those derivatives are given by the following expressions:
	Let us calculate the derivative of $A_z$ first
		\begin{align*}
			\hspace{1em}&\hspace{-1em}
			\partial_T\left(A_z(t,T)\right)\\&=
    \phi^3_z\left(\frac{\phi^1_z\phi^2_z{e}^{\phi^2_z\left(T-t\right)}}{\phi^1_z+\phi^2_z\left({e}^{\phi^1_z\left(T-t\right)}-1\right)}-\frac{\left(\phi^1_z\right)^2\phi^2_z{e}^{\phi^1_z\,\left(T-t\right)}{e}^{\phi^2_z\left(T-t\right)}}{{\left(\phi^1_z+\phi^2_z\left({e}^{\phi^1_z\left(T-t\right)}-1\right)\right)}^2}\right){\left(\frac{\phi^1_z{e}^{\phi^2_z\left(T-t\right)}}{\phi^1_z+\phi^2_z\left({e}^{\phi^1_z\left(T-t\right)}-1\right)}\right)}^{\phi^3_z-1}.
		\end{align*}
	Hence, we get
		\begin{align*}
			-\frac{\partial_T\left(A_x(t,T)\right)}{A_x(t,T)}=
			\frac{
				\phi^2_z \phi^3_z \left(\phi^1_z-\phi^2_z\right) \left({e}^{(T-t)\phi^1_z}-1\right)
			}{
				\phi^1_z+\phi^2_z\left({e}^{(T-t)\phi^1_z}-1\right)
			}.
		\end{align*}
	Now, we compute the derivative of $B_z$
		\begin{align*}
			\partial_T\left(B_z(t,T)\right)=
				\frac{
					\left(\phi^1_z\right)^2 {e}^{(T-t) \phi^1_z}
				}{
					{\left(\phi^1_z+\phi^2_z\left({e}^{(T-t) \phi^1_z}-1\right)\right)}^2
				}.
		\end{align*}

\section{Mean error of discount factors}\label{sec:meanErrDF}
\begin{table}[!ht]
\caption{Mean error of discount factors (DF) of our model with parameters given in \Cref{tab:risultati_cal} at \dateA and \dateB.}
	\centering
	\small
		\begin{tabular}{|cc}
			\hline
			Maturity (in years) & Mean Error of DF at \dateA\\
			\hline
			\dfMeanErrABody\\
			\hline
		\end{tabular}%
		\begin{tabular}{c|}
			\hline
			Mean Error of DF at \dateB\\
			\hline
			\dfMeanErrBData\\
			\hline
		\end{tabular}
	\label{tab:dfMeanErrAB}
\end{table}
\section{Market data}\label{sec:data}
\begin{table}[!ht]
	\caption{Market data containing the zero rate curve and zero coupon curve at \dateA.}
	\centering
	\small
	\marketDataA
	\label{tab:first_curve}
\end{table}

\begin{table}[!ht]
	\caption{Market data containing the zero rate curve and zero coupon curve at \dateB.}
	\centering
	\small
	\marketDataB
	\label{tab:second_curve}
\end{table}

\begin{table}[!ht]
	\caption{Market data containing the swaption prices at \dateA.}
	\centering
	\small
	\marketSwaptionA
	\label{tab:swapPrices_1}
\end{table}

\begin{table}[!ht]
	\caption{Market data containing the swaption prices at \dateB.}
	\centering
	\small
	\marketSwaptionB
	\label{tab:swapPrices_2}
\end{table}

\begin{table}[!ht]
	\caption{Market data containing the swaption strikes at \dateA.}
	\centering
	\small
	\strikeSwaptionA
	\label{tab:swapStrike_1}
\end{table}

\begin{table}[!ht]
	\caption{Market data containing the swaption strikes at \dateB.}
	\centering
	\small
	\strikeSwaptionB
	\label{tab:swapStrike_2}
\end{table}

\section*{Declarations}
\subsection*{Funding}
This project has received funding from the European Union’s Horizon 2020 research and innovation
programme under the Marie Sklodowska-Curie grant agreement No 813261 and is part of the ABC-EU-XVA project.
\subsection*{Conflicts of interests}

The authors have no relevant financial or non-financial interests to disclose.

\subsection*{Data availability}
All data generated or analysed during this study are included in this published article.
{
In particular the code to produce the numerical experiments is available at\\
\url{https://github.com/kevinkamm/CIR-}.
}
{\thispagestyle{scrheadings}
\newpage
\thispagestyle{scrheadings}\ihead{}
\singlespacing
\begin{footnotesize}
\bibliographystyle{acm}
\bibliography{literature.bib}
\end{footnotesize}
}
\end{document}